\documentclass[a4paper,11pt]{article}
\usepackage{pos}
\usepackage{subfigure} 

\newcommand{\mpi}{M_\pi}

\newcommand{\MSb}{\overline{\text{MS}}}
\DeclareMathOperator{\Tr}{Tr}

\begin{document}

\title{Flavor Diagonal Nucleon Charges}

\author*[a,b,c]{Sungwoo Park}
\author[a]{Tanmoy Bhattacharya}
\author[a]{Rajan Gupta}
\author[e]{Huey-Wen Lin}
\author[a,e]{Santanu Mondal}
\author[d]{Boram Yoon}
\author[e]{Rui Zhang}
\affiliation[a]{Theoretical Division T-2, Los Alamos National Laboratory, Los Alamos, NM 87545, USA}
\affiliation[b]{Center for Nonlinear Studies, Los Alamos National Laboratory, Los Alamos, NM 87545, USA}
\affiliation[c]{Thomas Jefferson National Accelerator Facility,
  12000 Jefferson Avenue, Newport News, VA 23606, USA}
\affiliation[d]{Computer, Computational and Statistical Science Division CCS-7, Los Alamos National Laboratory, Los Alamos, NM 87545, USA}
\affiliation[e]{Department of Physics and Astronomy, Michigan State University, East Lansing, MI 48824, USA}

\emailAdd{sungwoo@jlab.org}
\emailAdd{rajan@lanl.gov}
\emailAdd{tanmoy@lanl.gov}
\emailAdd{santanu@lanl.gov}
\emailAdd{boram@lanl.gov}
\emailAdd{hueywen@msu.edu}
\emailAdd{zhangr60@msu.edu}

\abstract{This talk provides an update on the calculation of matrix
  elements of flavor diagonal axial, scalar and tensor quark bilinear
  operators between the nucleon ground state. The simulations are done
  using Wilson-clover fermions on a sea of eight 2+1+1-flavor HISQ
  ensembles generated by the MILC collaboration. We discuss the signal
  in the connected and disconnected contributions, calculation of the
  renormalization constants and mixing in the RI-sMOM scheme, and
  control over the simultaneous chiral-continuum-finite-volume fit
  used to extract the final charges. }

\FullConference{%
 The 38th International Symposium on Lattice Field Theory, LATTICE2021
  26th-30th July, 2021
  Zoom/Gather@Massachusetts Institute of Technology
}

\maketitle

\section{Introduction}

Results for the matrix elements of flavor diagonal axial, scalar and
tensor quark bilinear operators between the nucleon ground state
provide a number of quantities of phenomenological interest. The axial
charges give contribution of each quark flavor to the spin of the
nucleon and the spin dependent interaction of dark matter with nuclear
targets; the tensor charges give the contribution of the quark
electric dipole moment (EDM) operator to the nucleon EDM and the
zeroth moment of transversity distribution of quarks in nucleons; and the flavor
diagonal scalar charges give the pion-nucleon sigma term, strangeness
content of the nucleon, the strength of the spin-independent coupling
of dark matter to nucleons, and enters in the search for BSM physics
such as in $\mu \to e$ conversion. This talk presents the status of
our calculations of these matrix elements on eight 
2+1+1-flavor HISQ ensembles generated by the MILC
collaboration~\cite{Bazavov:2012xda} using Wilson-clover valence
fermions with quark masses tuned to reproduce the sea $M_\pi$ and
$M_{s \bar s}$ values. The parameters of these ensembles are given in
Table~\ref{tab:hisq}. This set includes one physical $M_\pi\approx
138$ MeV ensemble (labeled as $a09m130$) at $a\approx 0.09$ fm and
$M_\pi L\approx 3.9$.

\begin{figure}[ht]
  \centering
  \begin{subfigure}[Nucleon charges]{
      \includegraphics[width=0.235\linewidth]{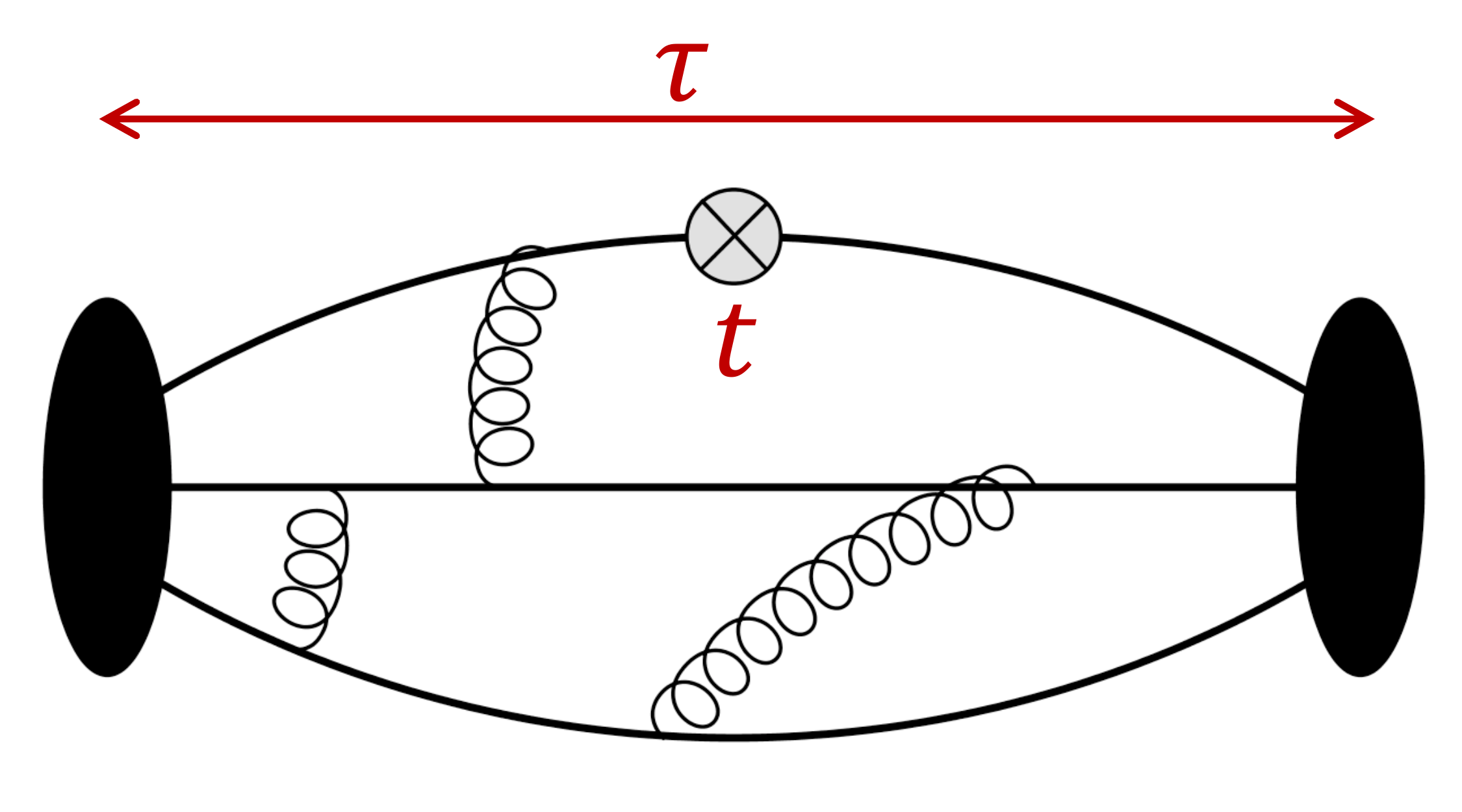}
      \includegraphics[width=0.235\linewidth]{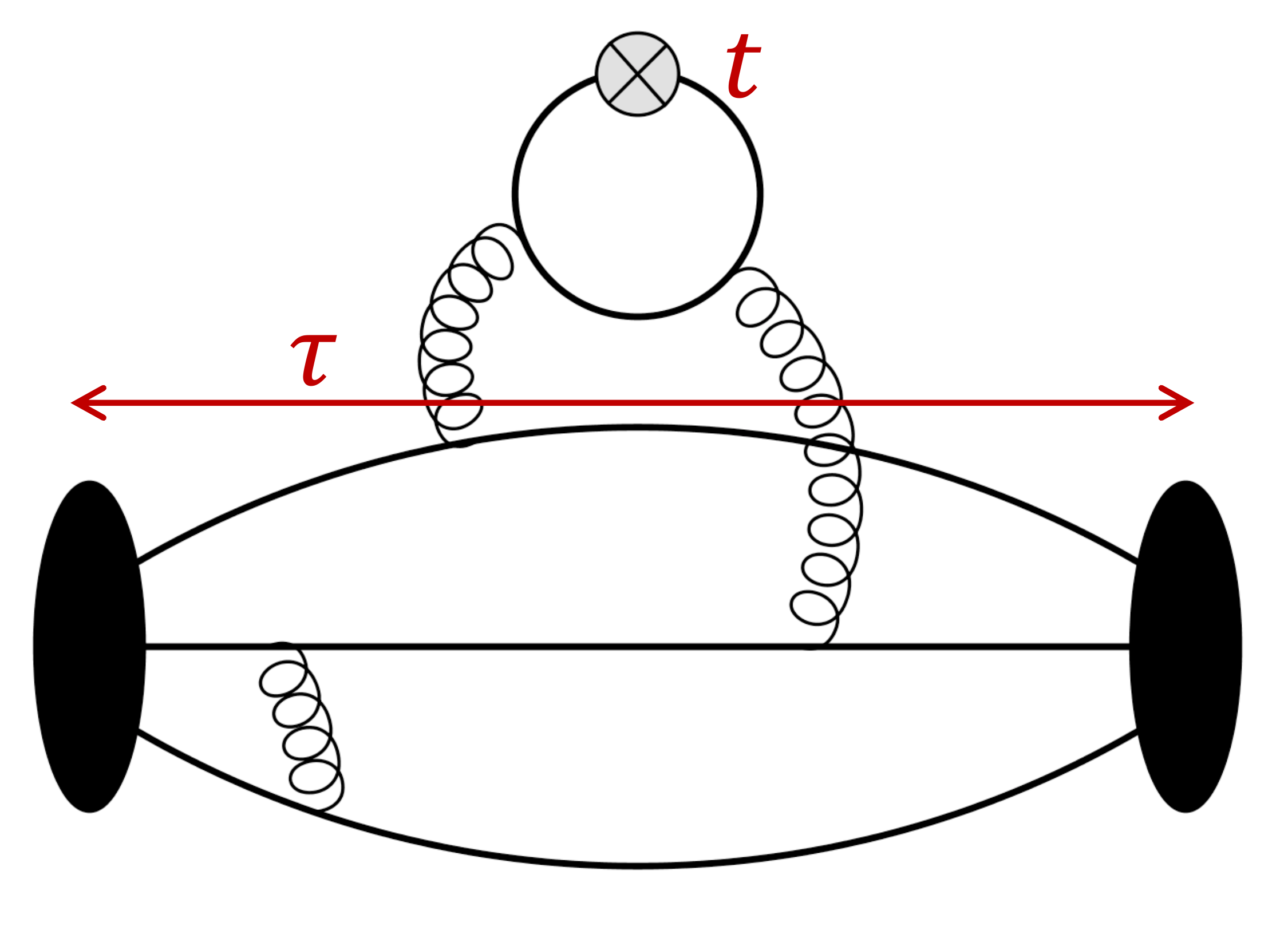}
      \label{fig:diag_g}
    }
  \end{subfigure}
  \begin{subfigure}[NPR]{
      \includegraphics[width=0.235\linewidth]{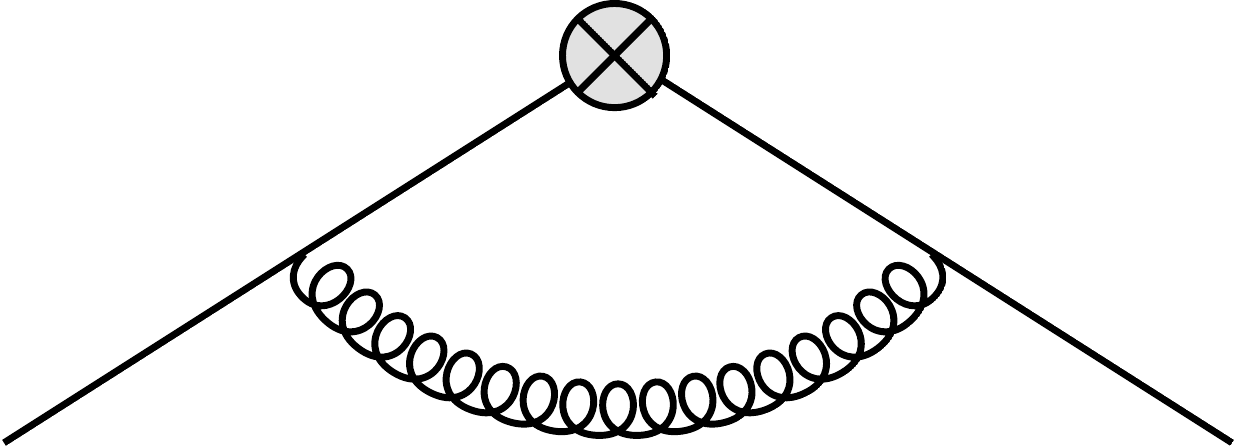}
      \includegraphics[width=0.235\linewidth]{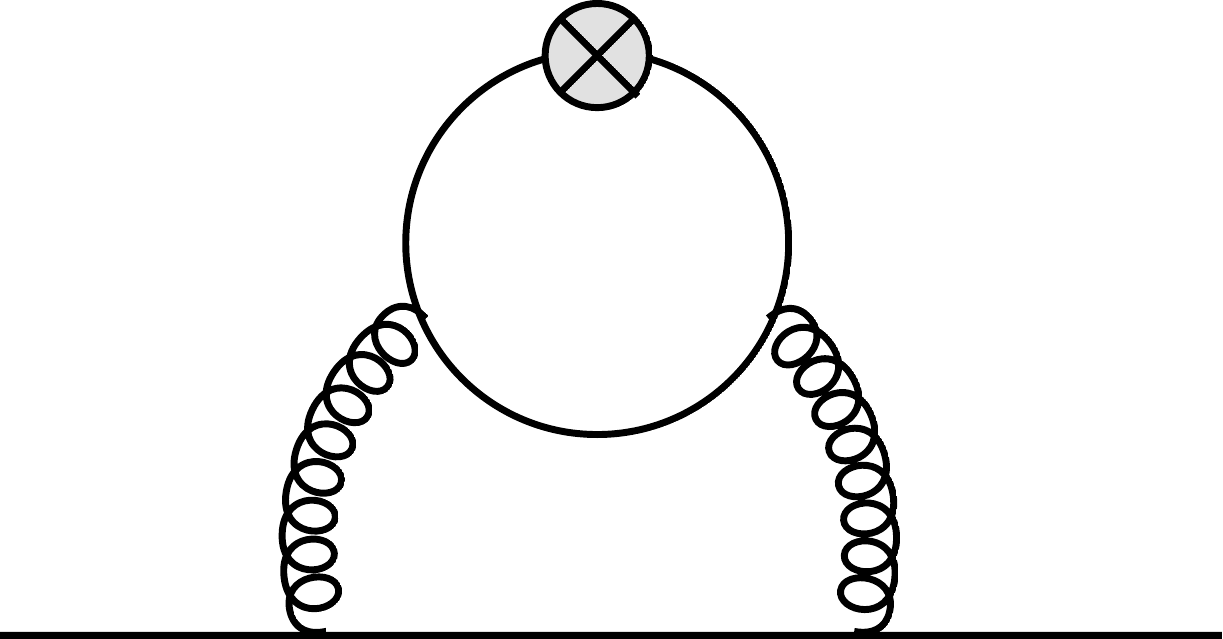}
      \label{fig:diag_npr}
    }
  \end{subfigure}
  \caption{The connected and disconnected diagrams calculated (i) for flavor diagonal nucleon charges,  and (ii) 
    non-perturbative renormalization in the RI-sMOM scheme using quark states in Landau gauge.} 
\end{figure}

All results presented here should be considered preliminary unless
otherwise stated. They are updates on results for $g_{A,T}^q$ in
Refs.~\cite{Gupta:2018qil,Gupta:2018lvp,Lin:2018obj}, and for $g_S^q$
in Ref.~\cite{Gupta:2021ahb}.  In addition to nucleon 2-point
functions, we calculate the connected~\cite{Gupta:2018qil} and
disconnected~\cite{Gupta:2018lvp,Lin:2018obj} contributions to 3-point
functions illustrated in the two left panels in Fig.~\ref{fig:diag_g},
and the analogous quark level diagrams in Landau gauge for calculating
the renormalization constants in the RI-sMOM scheme Fig.~\ref{fig:diag_npr}. 

\begin{table}[hb]
  \vspace{-0.5mm}
\center  
\resizebox{\textwidth}{!}{
\begin{tabular}{l|cccc|ccccccccccccccccc}
Ensemble ID    & $a$ (fm) & $M_\pi$ (MeV)    & $M_\pi L$  &  $L^3\times T$ & $N_\text{conf}^l$ & $N_\text{src}^l$ & $N_\text{conf}^s$ &
    $N_\text{src}^s$ & $N_\text{LP}/N_\text{HP}$\\\hline
$a15m310$    & 0.1510(20) & 320(5)      & 3.93    & $16^3\times 48$   & 1917 & 2000  & 1917 & 2000  & 50    \\
$a12m310$    & 0.1207(11) & 310(3)      & 4.55    & $24^3\times 64$   & 1013 & 10000 & 1013 & 8000  & 50    \\
$a12m220$    & 0.1184(10) & 228(2)      & 4.38    & $32^3\times 64$   & 958  & 11000 & 870  & 5000  & 30--50\\
$a09m310$    & 0.0888(8) &  313(3)      & 4.51    & $32^3\times 96$   & 1017 & 10000 & 889  & 6000  & 50    \\
$a09m220$    & 0.0872(7) &  226(2)      & 4.79    & $48^3\times 96$   & 712  & 8000  & 847  & 10000 & 30--50\\
$a09m130$    & 0.0871(6) &  138(1)      & 3.90    & $64^3\times 96$   & 1270 & 10000 & 541  & 10000 & 50    \\
$a06m310$    & 0.0582(4) &  320(2)      & 3.90    & $48^3\times 144$  & 808  & 12000 & 948  & 10000 & 50    \\
$a06m220$    & 0.0578(4) &  235(2)      & 4.41    & $64^3\times 144$  & 1001 & 10000 & 1002 & 10000 & 50    \\
\end{tabular}}
\vspace{-2mm}
\caption{Update from Refs.~\cite{Lin:2018obj,Gupta:2018lvp} of the
  statistics and $2+1+1$-flavor HISQ ensembles used for the
  calculation of disconnected contributions. Statistics for the connected are the same as in
  Ref.~\cite{Gupta:2018qil}.  $N_\text{conf}^{l,s}$ is the number of
  gauge configurations analyzed for light ($l$) and strange ($s$)
  flavors. $N_\text{src}^{l,s}$ is the number of random sources used
  per configurations, and $N_\text{LP}/N_\text{HP}$ is the ratio of
  low/high precision meausurements.  }
\label{tab:hisq}
\end{table}
  \vspace{-0.9mm}

\section{Details of 2-point and 3-point function analysis}

Details of the calculations of the quark propagators on HYP smeared
lattices using Wuppertal smearing are given in
Ref.~\cite{Gupta:2018qil}.  The nucleon interpolating operator used
both at the source and the sink in all calculations is $\mathcal{N}(x)
= \epsilon^{abc} \left[ {q_1^a}^T(x) C \gamma_5 \frac{(1 \pm
    \gamma_4)}{2} q_2^b(x) \right] q_1^c(x) $.  All 3-point functions
are calculated with zero momentum operator insertion and the nucleon
state at the sink is also projected to zero momentum. From these we get 
forward matrix elements from which charges are obtained. 

The nucleon spectrum is obtained from the 2pt function, with spin
projection $(1+\gamma_4)/2$, fit using the
spectral decomposition truncated at four states, 
  \begin{align}
    C^\textrm{2pt}(\tau)=\sum_{i=0} | \mathcal{A}_i|^2 e^{-M_i \tau}. 
  \end{align}
We carry out two types of analyses: The ``standard'' fit uses wide
priors for all the excited-state amplitudes, $\mathcal{A}_i$, and
masses, $M_i$, i.e., the priors are only used to stabilize the
fits. This is called the $\{4\}$ or ``standard'' strategy. (ii) A narrow prior for
$M_1$ with central value given by the non-interacting energy of the
lowest allowed $N \pi$ or $N \pi \pi$ state is used.  This is called
the $\{4^{N\pi}\}$ strategy. The resulting values of $\mathcal{A}_0$
and the $M_i$ are used as inputs in the analysis of the 3-point
functions. The mass gap, $M_1-M_0$, in the 2 analyses is
significantly different, however the augmented $\chi^2$ minimized in the fits is
essentially the same. Thus, the two strategies are not distinguished on the 
basis of our fits and we examine the senstivity of the results for the charges 
to the two $M_1$.

The nucleon 3pt function at zero momentum, in our setup, is given by 
  \begin{align}
    C^{\text{3pt}}_\Gamma(t;\tau)=\Tr [ P_\Gamma \langle 0|\mathcal{N}(\tau) 
    O_\Gamma(t,{\textbf q=0}) \bar{\mathcal{N}}(0,{\textbf p=0})|0 \rangle ] \,, 
    \label{eq:3pt}
  \end{align}
where the operator $O_\Gamma^q= \bar q \Gamma q$, $q \in \{u,d,s\}$, and
$P_\Gamma = (1+\gamma_4)(1+i \gamma_5\gamma_3)/2$ is the spin
projection used for forward propagating nucleons. This
$P_\Gamma$ singles out direction ``3'', while ``1'' and ``2'' are
equivalent under the cubic rotational symmetry.  Flavor diagonal 3pt
functions require the sum of connected (conn) and disconnected (disc)
contributions illustrated in Fig.~\ref{fig:diag_g}:
  \begin{align}
    C^{\text{3pt}}_\Gamma(t;\tau)&=C^{\text{conn}}_\Gamma(t;\tau) + C^{\text{disc}}_\Gamma(t;\tau). 
  \end{align}
For the scalar case, the disconnected contribution is calculated using
the vacuum subtracted operator $O_S^q - \langle O_S^q \rangle$.  The
calculation of the quark loop with zero-momentum operator insertion is
estimated stocastically using $Z_4$ random noise sources as explained in
Ref.~\cite{Bhattacharya:2015wna}.

We analyze the zero-momentum nucleon 3pt function (sum of connected and disconnected diagram contributions) 
using the spectral decomposition 
  \begin{align}
    C^{\textrm{3pt}}_\Gamma (\tau;t)=\sum_{i,j=0} \mathcal{A}_i
    \mathcal{A}_j^\ast \langle i |O_\Gamma| j \rangle e^{-M_i t - M_j (t- \tau)},
\label{eq:3pt-sd}
  \end{align}
and obtain the bare charges, $g_\Gamma^{q;\text{bare}}$, from the
ground state matrix elements $\langle 0 | O_\Gamma | 0 \rangle$. The
challenge is removing excited state contributions (ESC) which are
observed to be large at source-sink separation $\tau\approx 1.4$ fm,
the typical distance beyond which the signal degrades due to
the $e^{(M_N-3/2M_\pi)\tau}$ increase in noise.  With the current
statistics, we are only able to keep one excited state in the analysis
using Eq.~\eqref{eq:3pt-sd} and fits leaving $M_1$ a free parameter
are not stable.  Thus we take $M_1$ from fits to the 2-point function,
i.e., we analyze the data with the two strategies, $\{4\}$ and
$\{4^{N\pi}\}$.
  
A challenge to distinguishing between $\{4\}$ and $\{4^{N\pi}\}$
strategies is that the difference in the corresponding $M_1$ becomes
significant only for $M_\pi \lesssim 200$~MeV, which in our setup
means only in the $a091m130$ ensemble. Previous work shows that the
difference in axial and tensor charges, $g_{A,T}$, is
small~\cite{Gupta:2018qil}.  For the isoscalar scalar charge
$g_S^{u+d}$, $\chi$PT suggests a large contribution from $N\pi$ and
$N\pi\pi$ states leading to a large difference in the value of the
pion-nucleon sigma term as explained in
Ref.~\cite{Gupta:2021ahb}. To understand and quantify these differences 
we do the full analysis with both strategies.\looseness-1

Note that in our previous works~\cite{Lin:2018obj,Gupta:2018lvp}, the
fits to remove ESC in $C^{\text{conn}}_\Gamma(t;\tau)$ and
$C^{\text{disc}}_\Gamma(t;\tau)$ were done separately, as was the
chiral-continuum (CC) extrapolations of $g_\Gamma^{q,\text{disc}}$ and
$g_\Gamma^{q,\text{conn}}$. This introduces an unquantified systematic~\cite{Lin:2018obj} 
that has now been removed by
fitting to $C^{\text{conn}}_\Gamma(t;\tau) +
C^{\text{disc}}_\Gamma(t;\tau)$ and extrapolating $g_\Gamma^{q}$.

\section{Renormalization}

We have now explicitly evaluated the $3\times 3$ flavor mixing
matrices in $\MSb$ scheme at $2$ GeV to get the renormalized flavor
diagonal axial, scalar and tensor charges.  The corrections are small
for axial and tensor charges as anticipated in
\cite{Lin:2018obj,Gupta:2018lvp}, but significant for the scalar
operators. For example, $g_S^s$ gets about $6\sim20$\% correction to the diagonal term
$Z_{S}^{s,s}g_S^{s}$ from
$Z_S^{s,u+d}g_S^{u+d,\text{bare}}$~\cite{Park:2020axe}.

For $N_f=2+1$-flavor theory, the mixing matrix for bilinear operators
$O_\Gamma^f$, $f\in \{u-d,u+d,s\}$ is \looseness-1
\begin{align}
  Z_\Gamma^\text{RI-sMOM}=
  \begin{pmatrix}
    Z_\Gamma^{u-d,u-d} & 0 & 0 \\
    0 & Z_\Gamma^{u+d,u+d} & Z_\Gamma^{u+d,s}\\
    0 & Z_\Gamma^{s,u+d} & Z_\Gamma^{ss} \\
  \end{pmatrix}
  =
  \begin{pmatrix}
    c_l & 0 & 0 \\
    0 & c_l-2d_{ll} & -2d_{sl} \\
    0 & -d_{ls} & c_s-d_{ss}\\
  \end{pmatrix}^{-1}
  \label{eq:Z_RI}
\end{align}
where $c_f$ and $d_{ff'}$ are the projected amputated Green's function
for the connected and disconnected contributions (Fig.~\ref{fig:diag_npr})
respectively. They are defined as follows,
\begin{align}
  c_f^\Gamma \equiv \frac{1}{Z_\psi^f}\text{Tr}[P^\text{RI-sMOM}_\Gamma \langle f |
    O_\Gamma^f | f \rangle_\text{conn} ] \,, \\
  d_{ff'}^\Gamma \equiv
  \frac{-1}{Z_\psi^f}\text{Tr}[P^\text{RI-sMOM}_\Gamma \langle f |
    O_\Gamma^{f'} | f \rangle_\text{disc} ] \,, 
\end{align}
with  $Z_\psi^f$ the wave function renormalization,
$P^\text{RI-sMOM}_\Gamma$ the projector for the RI-sMOM scheme, and 
$\langle f | O_\Gamma^f | f \rangle$ the 
amputated Green's functions. The quark loop is again estimated
stochastically. 
From Eq.~\eqref{eq:Z_RI}, we can calculate each $Z_\Gamma^\text{RI-sMOM}(q^2)$
from the associated 2 connected ($c_l^\Gamma$, $c_s^\Gamma$) and 4 disconnected ($d_{ll}^\Gamma$,
$d_{ls}^\Gamma$, $d_{sl}^\Gamma$, $d_{ss}^\Gamma$) projected amputated Green's
functions. Here $q^2$ is the momentum flowing in all three legs and defines the RI-sMOM scale.
The matrix $Z_\Gamma^\text{RI-sMOM}(q^2)$ is roughly diagonal since $c_l^\Gamma \sim
O(1)$ but $d_{ff'}^\Gamma$ are a few percent at $|q| \sim 2$~GeV for the 
scalar operator, and smaller still for the axial and tensor operators. \looseness-1

The four steps to get $Z_\Gamma^{\MSb}(\mu=2~\text{GeV})$ are: (i) from the 2
$c_f^\Gamma$ and 4 $d_{ff'}^\Gamma$, we calculate the full $3\times 3$ matrix
$Z_\Gamma^\text{RI-sMOM}(q^2)$ for various $q^2$ using
Eq.~\eqref{eq:Z_RI}. (ii) Perform horizontal matching to $\MSb$, 
$Z_\Gamma^{\MSb}(\mu)=C_\Gamma^{RI\to   \MSb}(\mu)Z_\Gamma^\text{RI-sMOM}(|q|)$, 
at scale $\mu = |q|$ using perturbation theory for $C_\Gamma^{RI\to   \MSb}(\mu)$. 
(iii) Perturbative running in $\MSb$ to fixed scale $ 2$GeV,
$Z_\Gamma^{\MSb}(2GeV;\mu)=C_\Gamma^{\MSb}(2GeV,\mu)Z_\Gamma^{\MSb}(\mu)$. (iv) Remove
dependence on $\mu^2 = q^2$ artifacts using the fit 
$Z_\Gamma^{\MSb}(2GeV;\mu)= Z_\Gamma^{\MSb}(2GeV) + c_1 \mu^2 + c_2 \mu^4$.

\section{Results for $g_{A,S,T}$}

\begin{figure}
  \center
  \begin{subfigure}[$g_{A}^{u}$]{\includegraphics[width=0.235\linewidth]{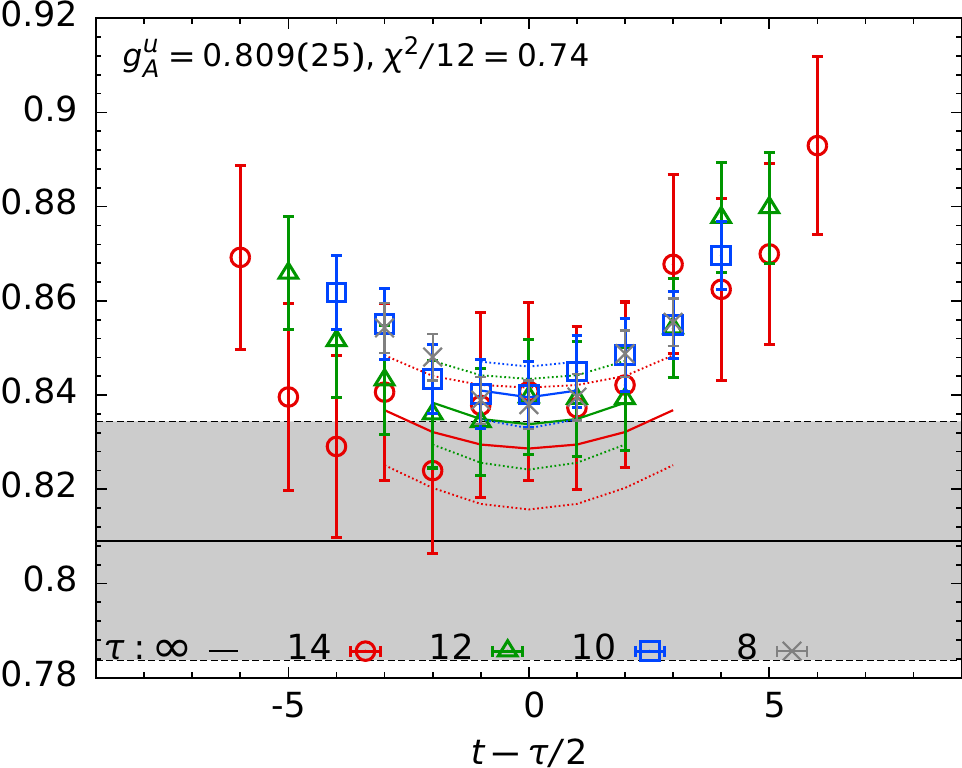}}\end{subfigure}
  \begin{subfigure}[$g_{A}^{d}$]{\includegraphics[width=0.235\linewidth]{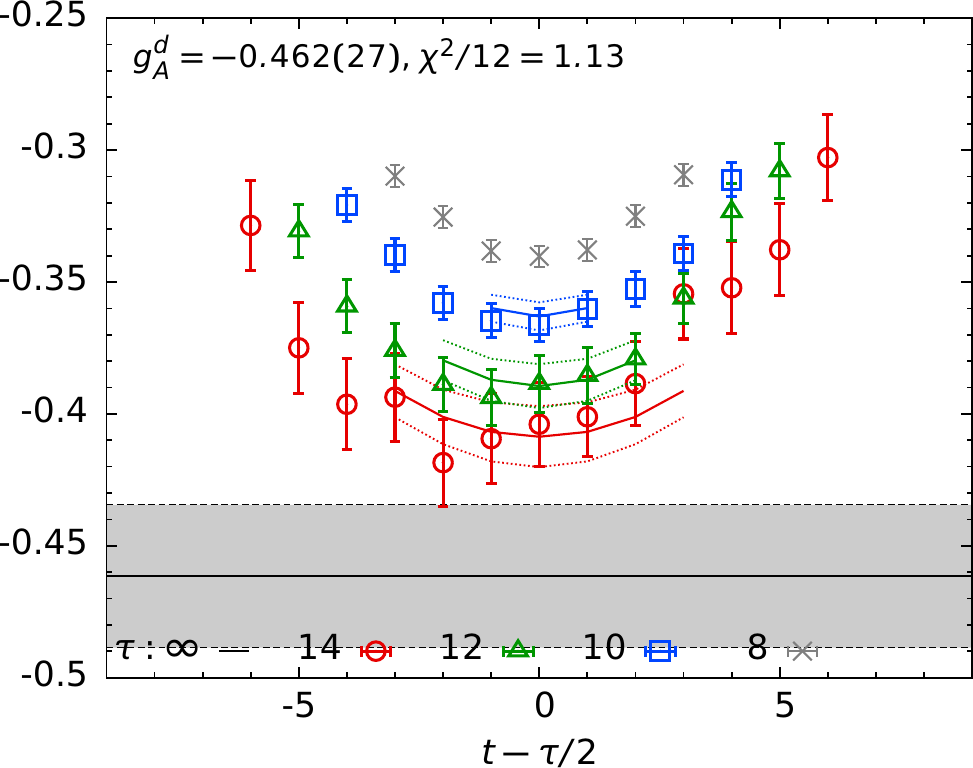}}\end{subfigure}
  \begin{subfigure}[$g_{A}^{u+d}$]{\includegraphics[width=0.235\linewidth]{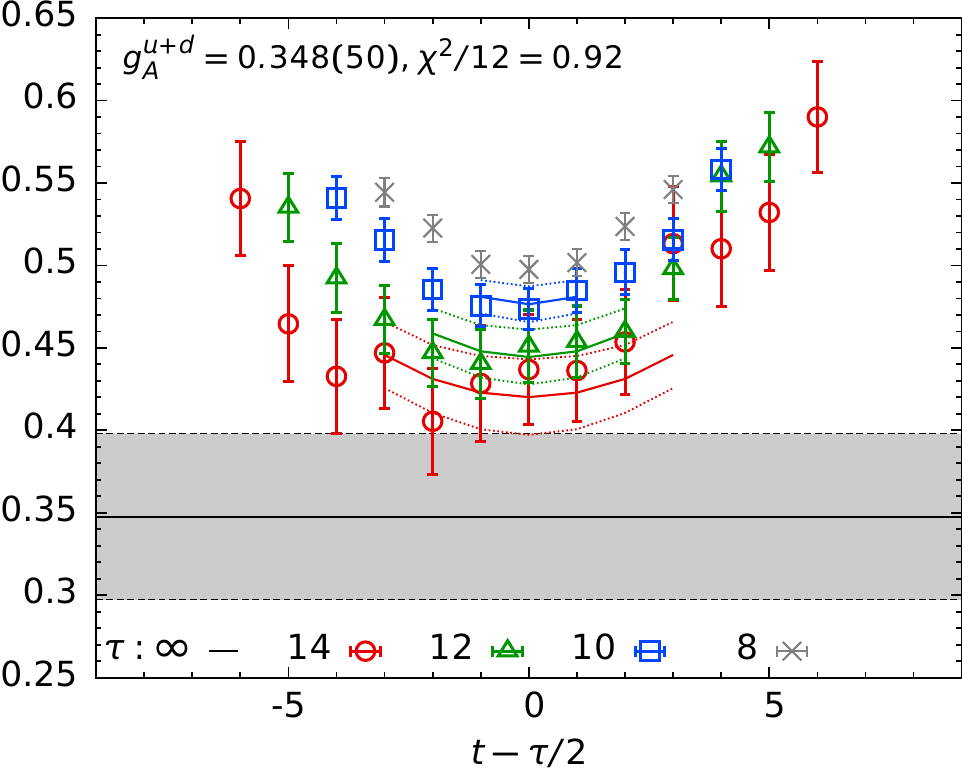}}\end{subfigure}
  \begin{subfigure}[$g_{A}^{s}$]{\includegraphics[width=0.235\linewidth]{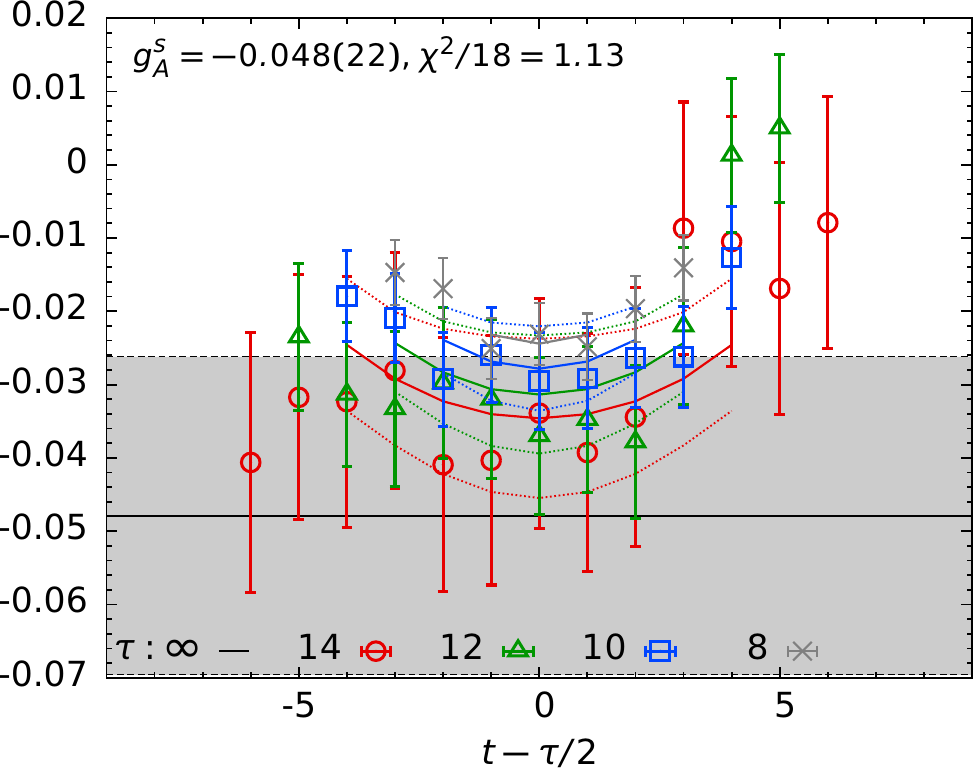}}\end{subfigure}

  \begin{subfigure}[$g_{T}^{u}$]{\includegraphics[width=0.235\linewidth]{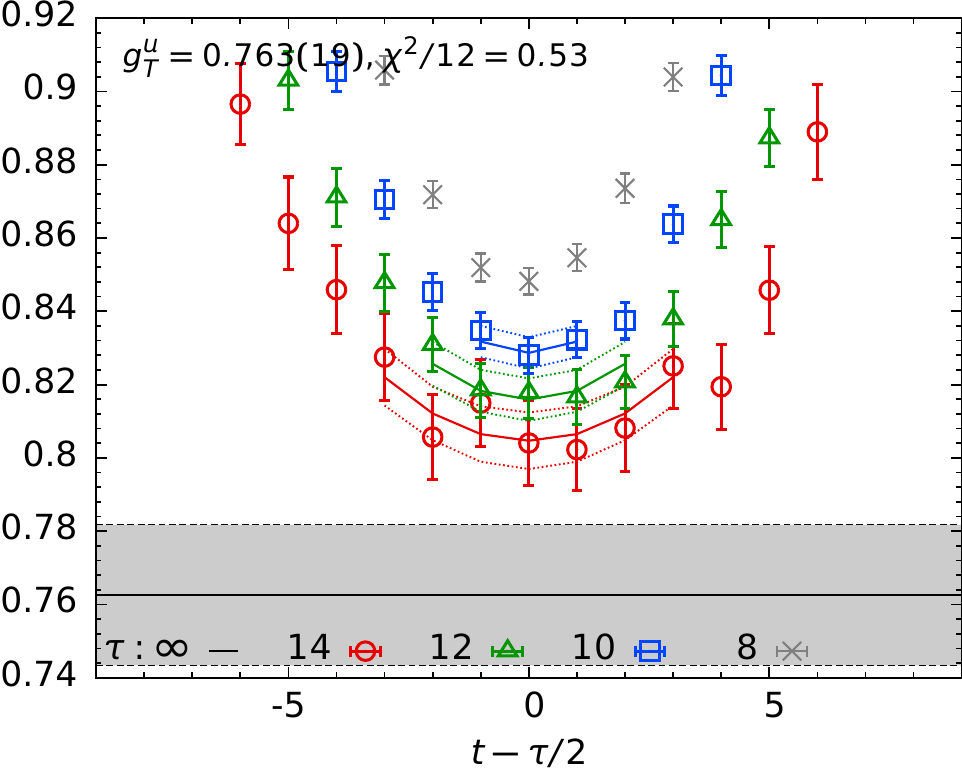}}\end{subfigure}
  \begin{subfigure}[$g_{T}^{d}$]{\includegraphics[width=0.235\linewidth]{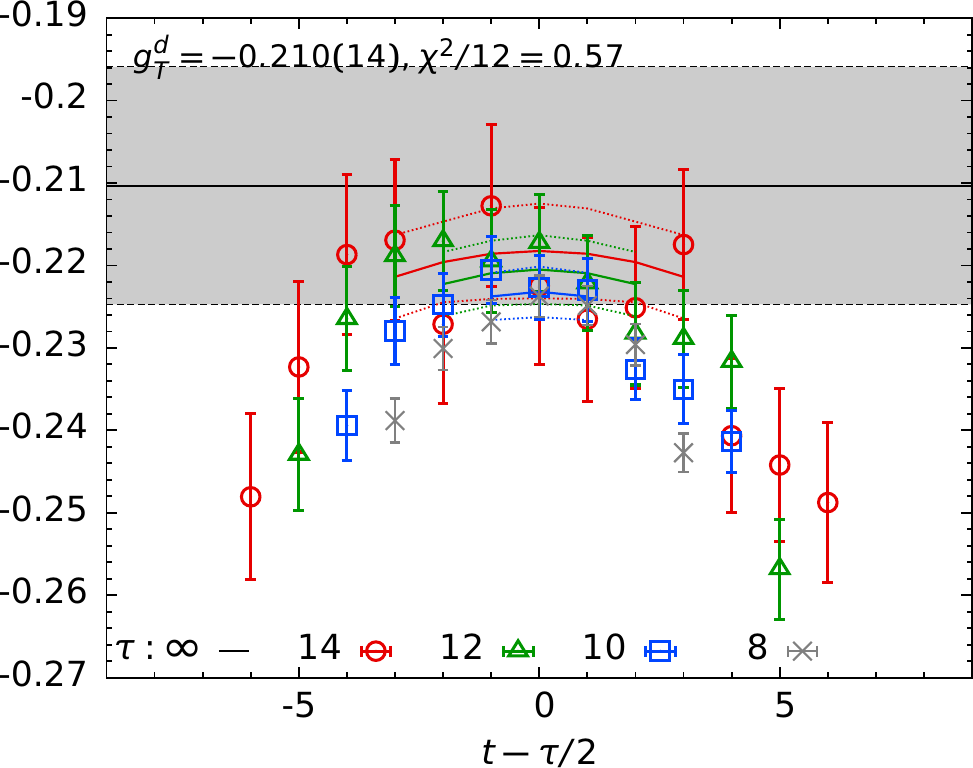}}\end{subfigure}
  \begin{subfigure}[$g_{T}^{u+d}$]{\includegraphics[width=0.235\linewidth]{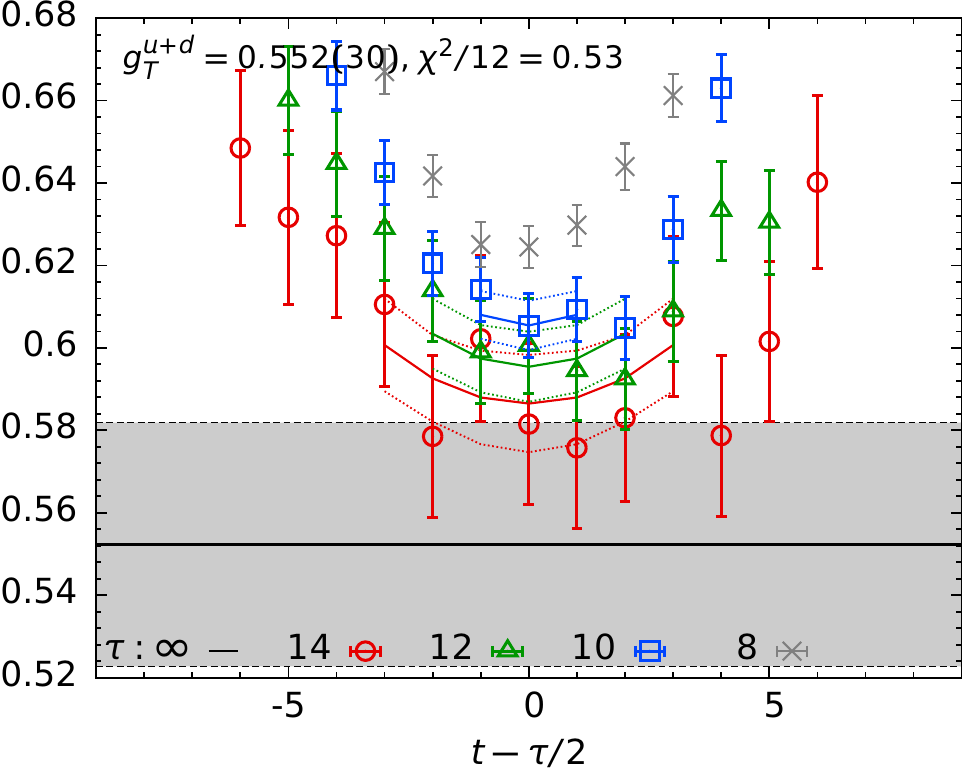}}\end{subfigure}
  \begin{subfigure}[$g_{T}^{s}$]{\includegraphics[width=0.235\linewidth]{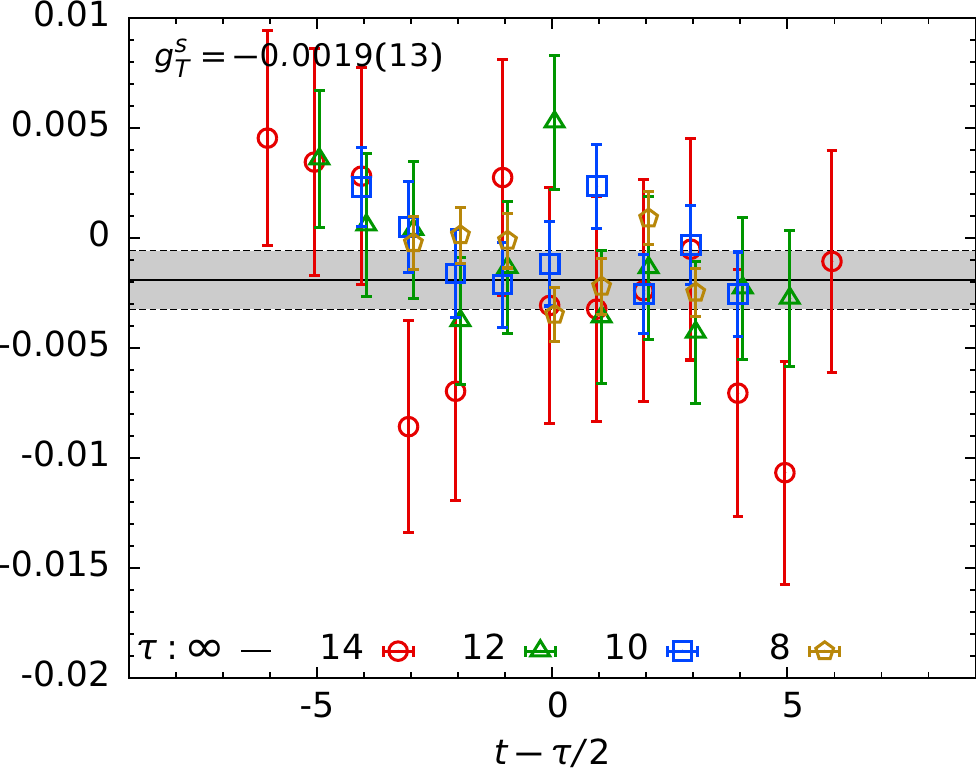}\label{fig:gT_s}}\end{subfigure}

  \caption{Results of bare $g_A$ (top) and $g_T$ (bottom) from the
    fits to the sum of the connected and disconnected data plotted
    versus $(t-\tau/2)/a$ for the physical $\mpi$ ensemble
    $a09m130$. Result of the fit is shown by lines of the same color
    as the data for various $\tau/a$ listed in the label, and the
    $\tau\to\infty$ is given by the gray band.}
  \label{fig:gAgT}
\end{figure}

\begin{figure}[b]  
  \center

  \begin{subfigure}[$g_{S}^{u}$]{\includegraphics[width=0.235\linewidth]{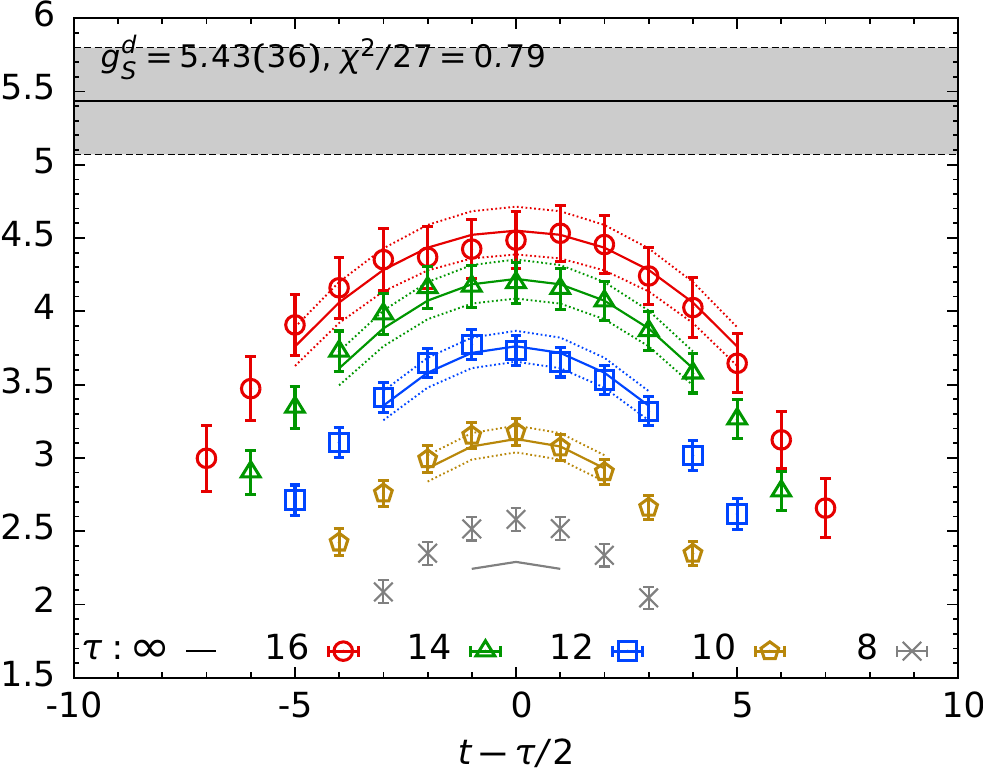}}\end{subfigure}
  \begin{subfigure}[$g_{S}^{d}$]{\includegraphics[width=0.235\linewidth]{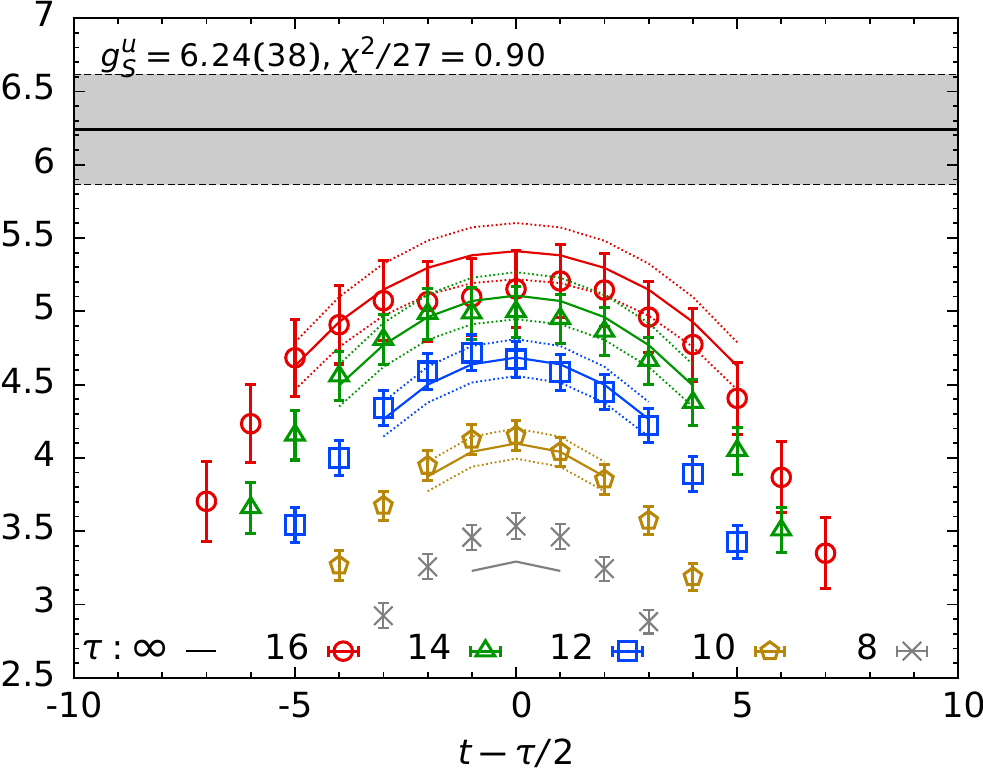}}\end{subfigure}
  \begin{subfigure}[$g_{S}^{u+d}$]{\includegraphics[width=0.235\linewidth]{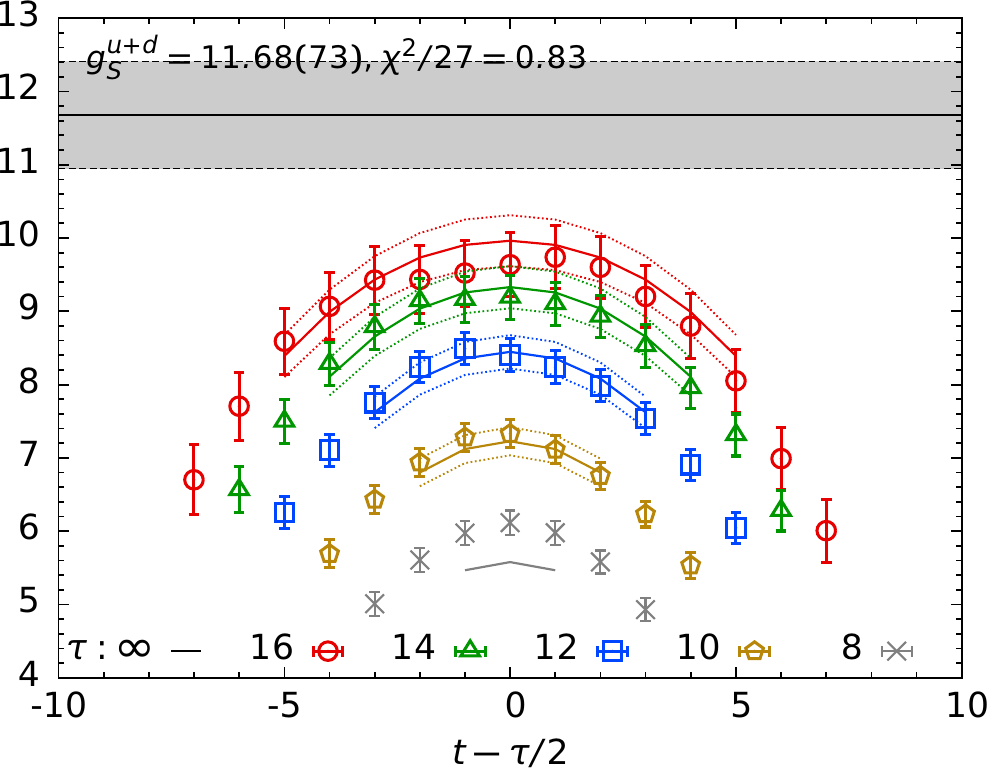}}\end{subfigure}
  \begin{subfigure}[$g_{S}^{s}$]{\includegraphics[width=0.235\linewidth]{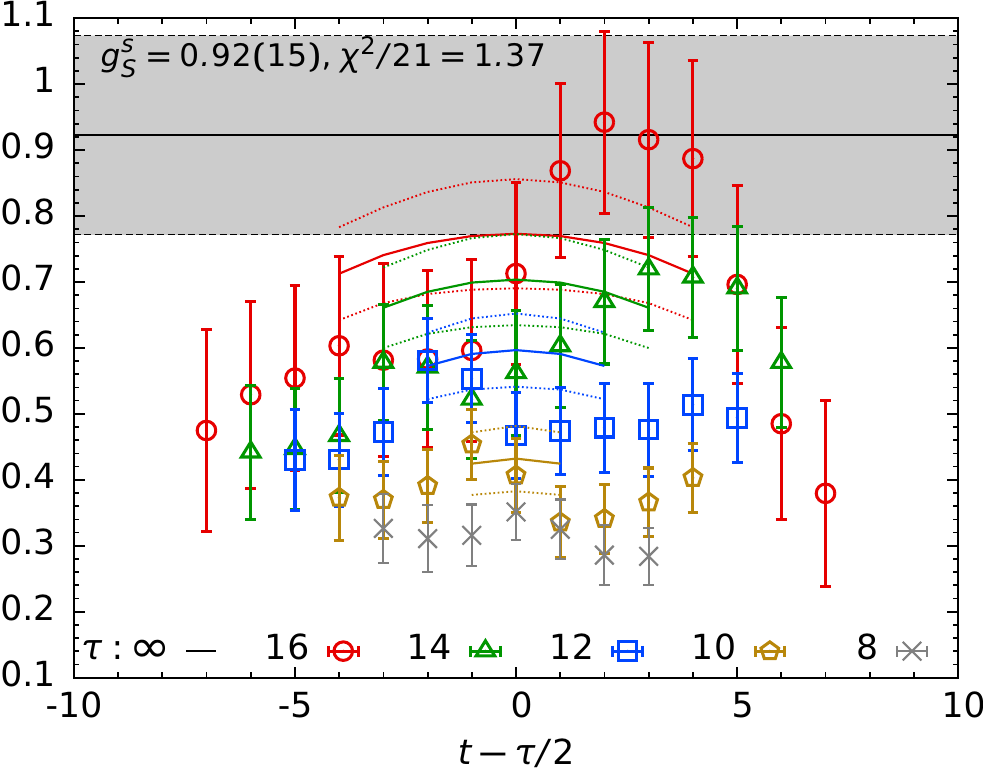}}\end{subfigure}

  \begin{subfigure}[$g_{S}^{u}$ with $N\pi$]{\includegraphics[width=0.235\linewidth]{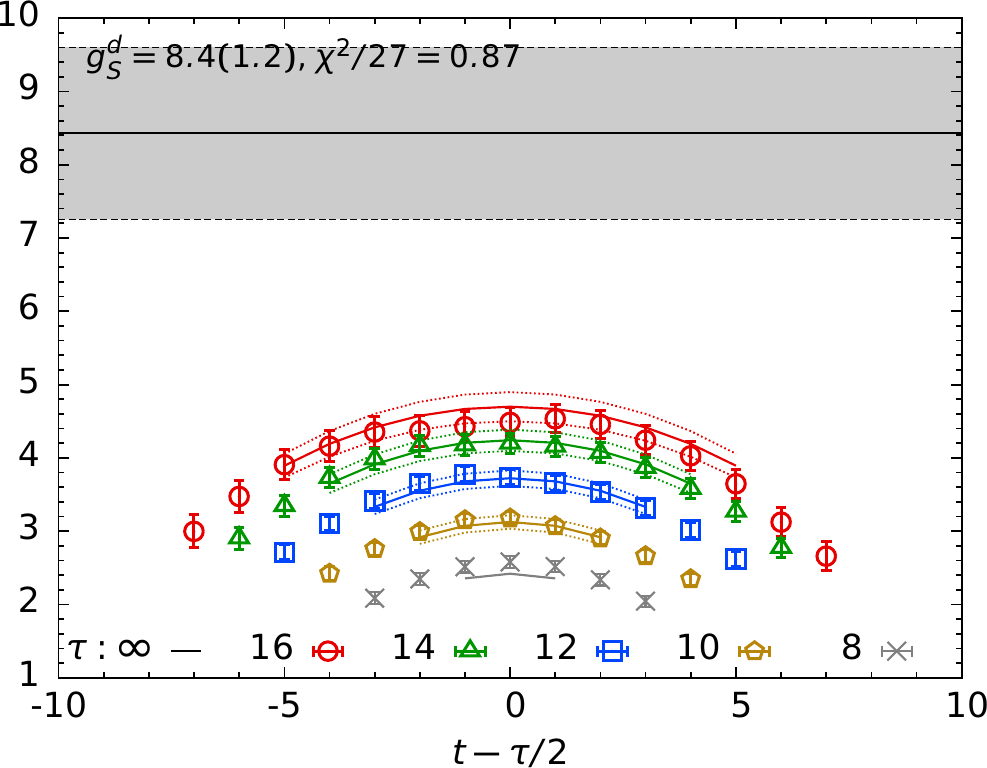}}\end{subfigure}
  \begin{subfigure}[$g_{S}^{d}$ with $N\pi$]{\includegraphics[width=0.235\linewidth]{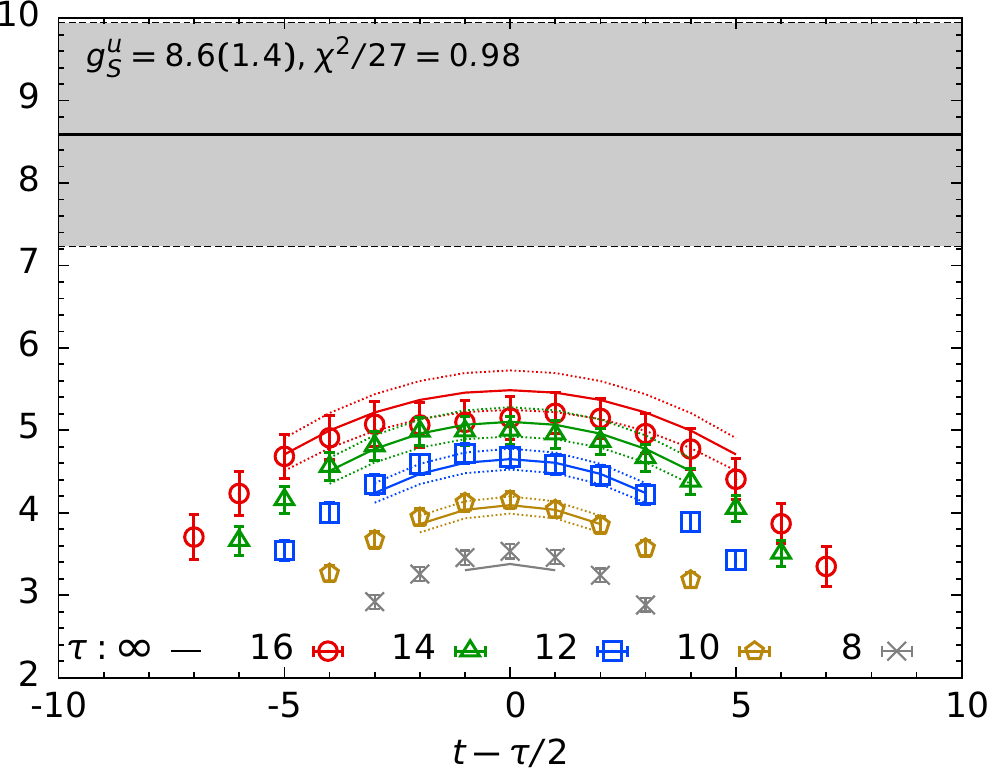}}\end{subfigure}
  \begin{subfigure}[$g_{S}^{u+d}$ with $N\pi$]{\includegraphics[width=0.235\linewidth]{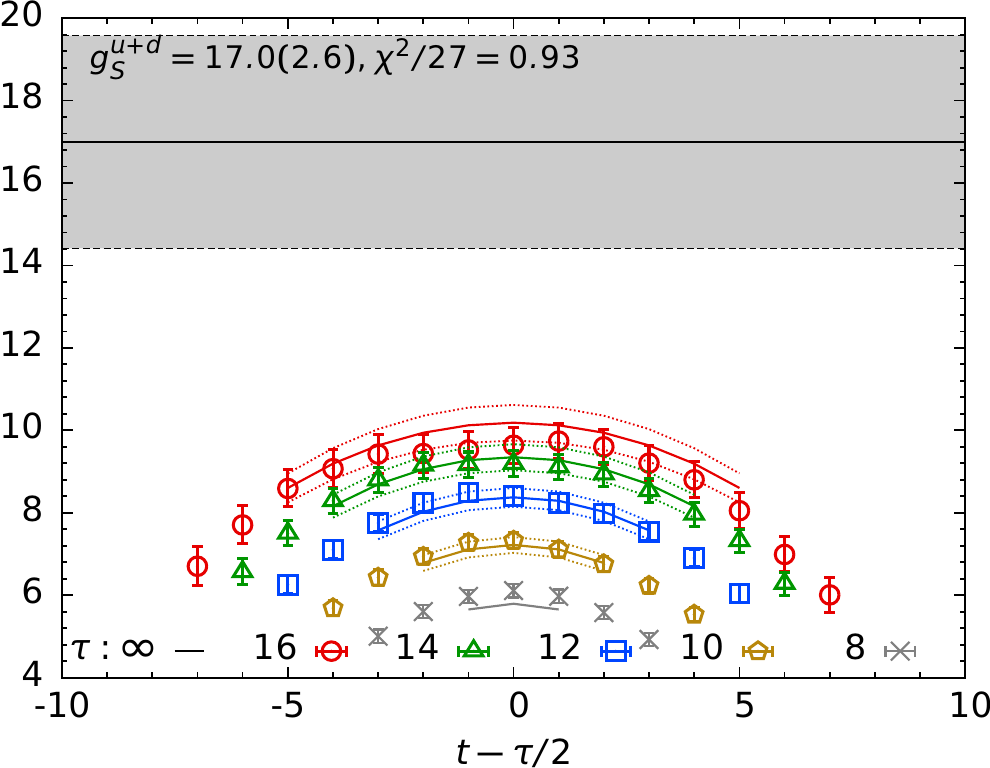}}\end{subfigure}
  \begin{subfigure}[$g_{S}^{s}$ with $N\pi$]{\includegraphics[width=0.235\linewidth]{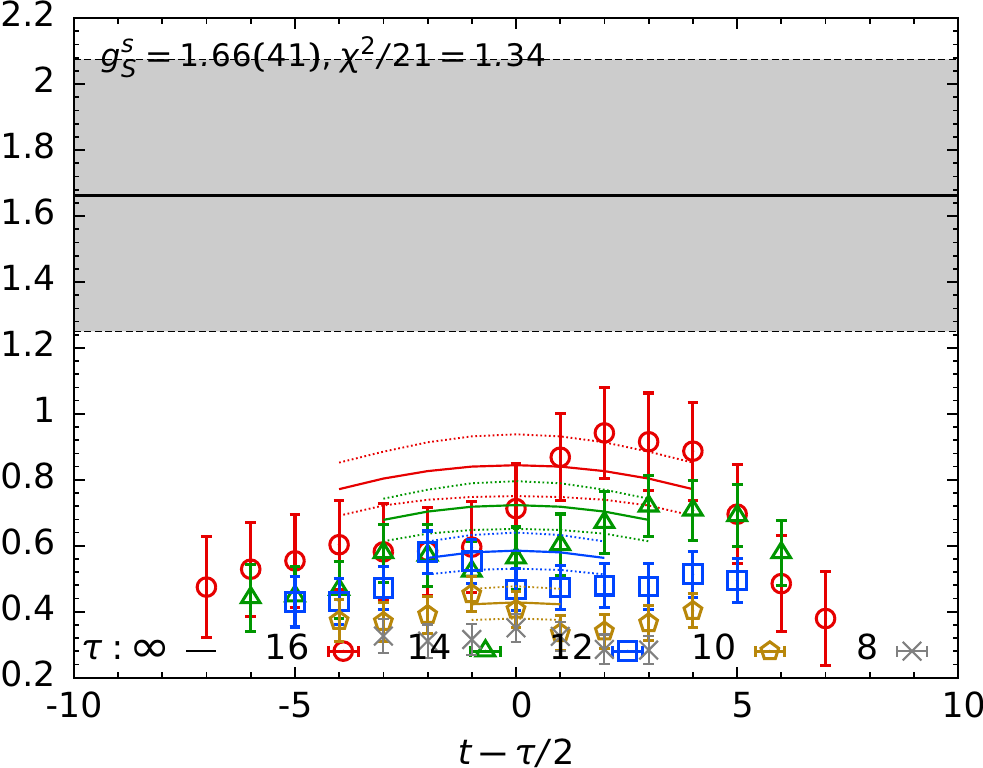}
  \label{fig:gS_s}}\end{subfigure}
  \vspace{-0.13in}
  \caption{Data for bare $g_S$ from the $a09m130$ ensemble (sum of connected
    and disconnected contributions) and fits using (i) the ``standard'' (top)
    and (ii) the ``$N\pi$'' (bottom) strategies. The rest is same as
    in Fig.~\ref{fig:gAgT}}
  \label{fig:gS}
  \vspace{-0.18in}
\end{figure}

\begin{figure}[p] 
  \centering
  \begin{subfigure}[$g_{A}^{u}$]{
      \includegraphics[width=0.235\linewidth]{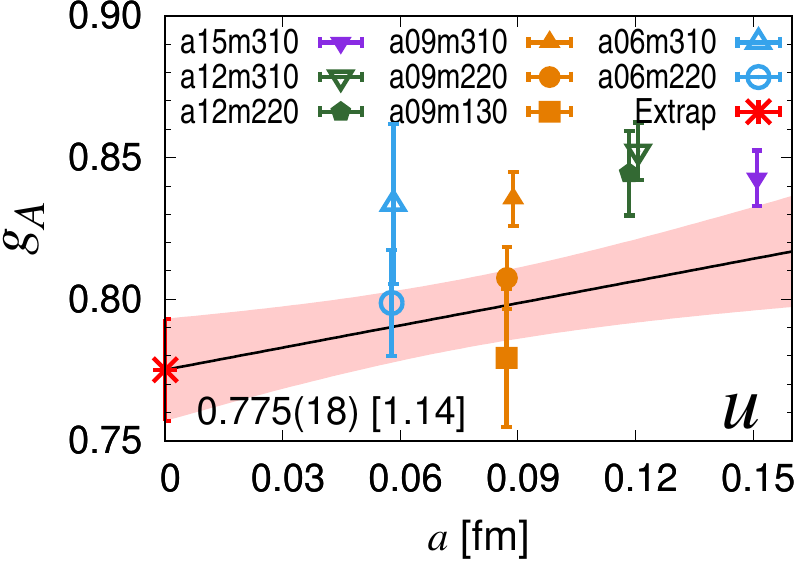}
      \includegraphics[width=0.235\linewidth]{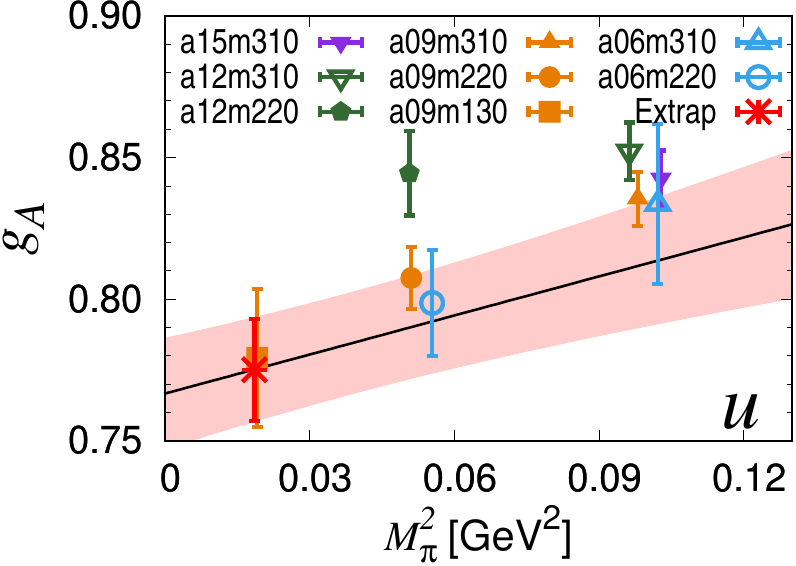}
  }\end{subfigure}
  \begin{subfigure}[$g_{A}^{d}$]{
      \includegraphics[width=0.235\linewidth]{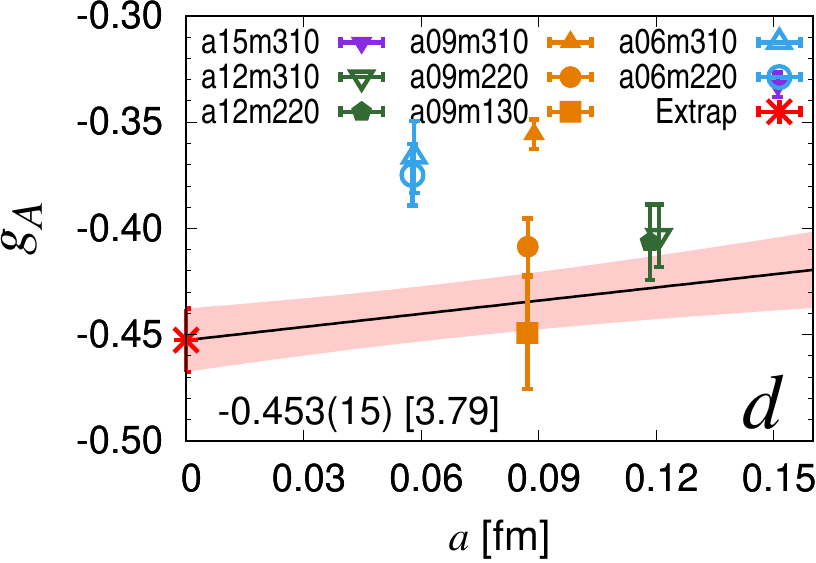}
      \includegraphics[width=0.235\linewidth]{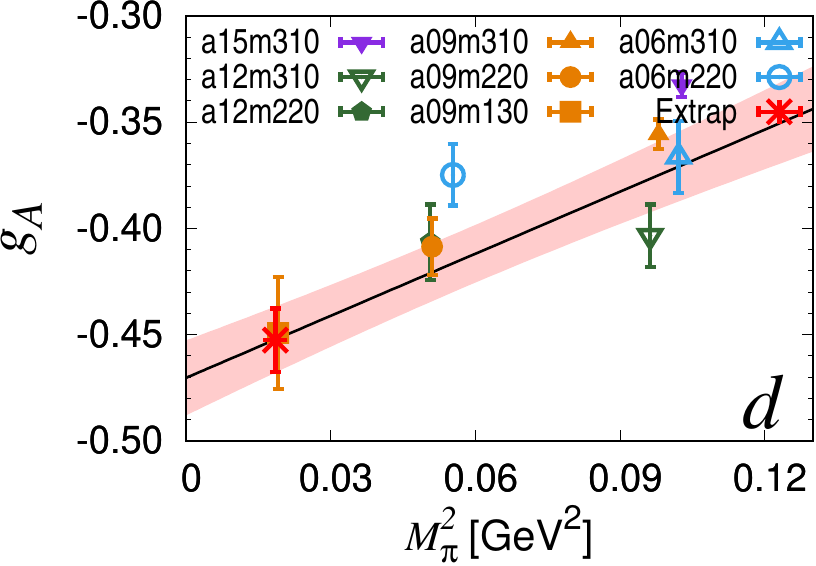}
  }\end{subfigure}
  \begin{subfigure}[$g_{A}^{u+d}$]{
      \includegraphics[width=0.235\linewidth]{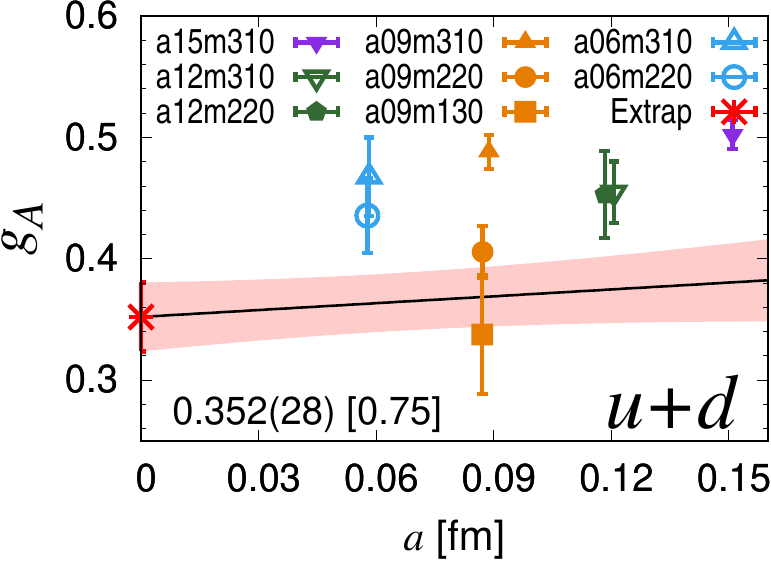}
      \includegraphics[width=0.235\linewidth]{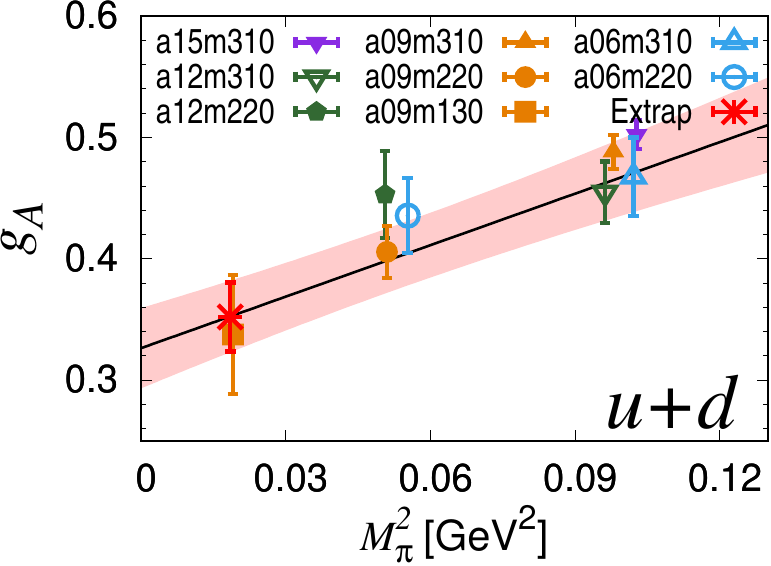}
  }\end{subfigure}
  \begin{subfigure}[$g_{A}^{s}$]{
      \includegraphics[width=0.235\linewidth]{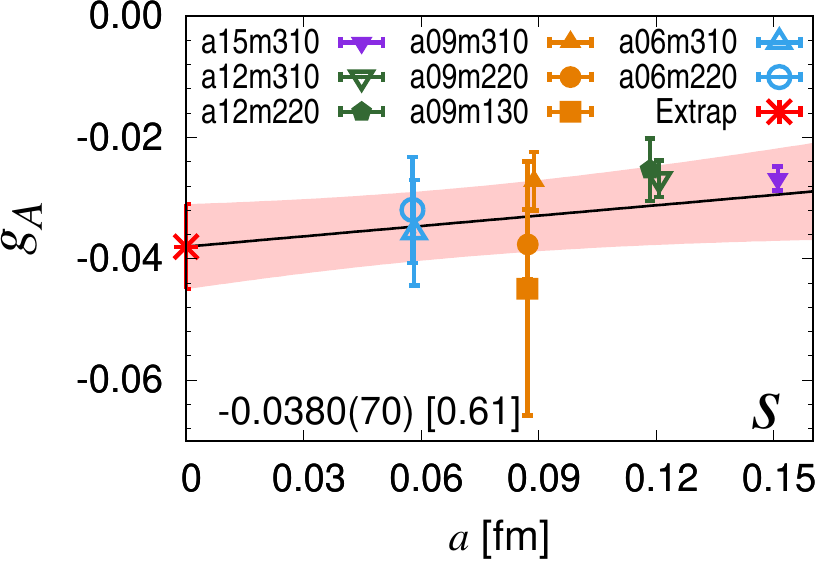}
      \includegraphics[width=0.235\linewidth]{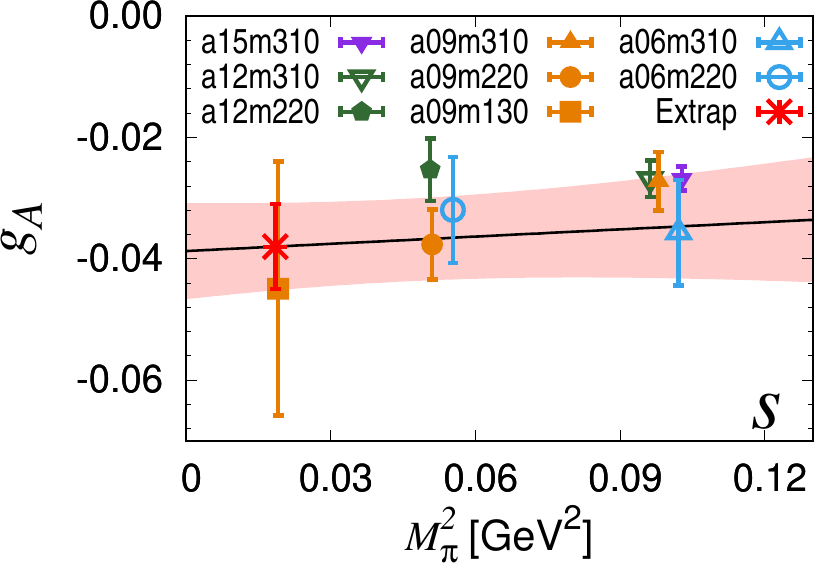}
  }\end{subfigure}
  \vspace{-0.1in}
  \caption{Chiral-continuum fits to $g_{A}$ using the ansatz $d_0+d_a a+ d_2 M_\pi^2$}
  \label{fig:CC_gA}
\end{figure}

\begin{figure}[p] 
  \centering
  \begin{subfigure}[$g_{T}^{u}$]{
      \includegraphics[width=0.235\linewidth]{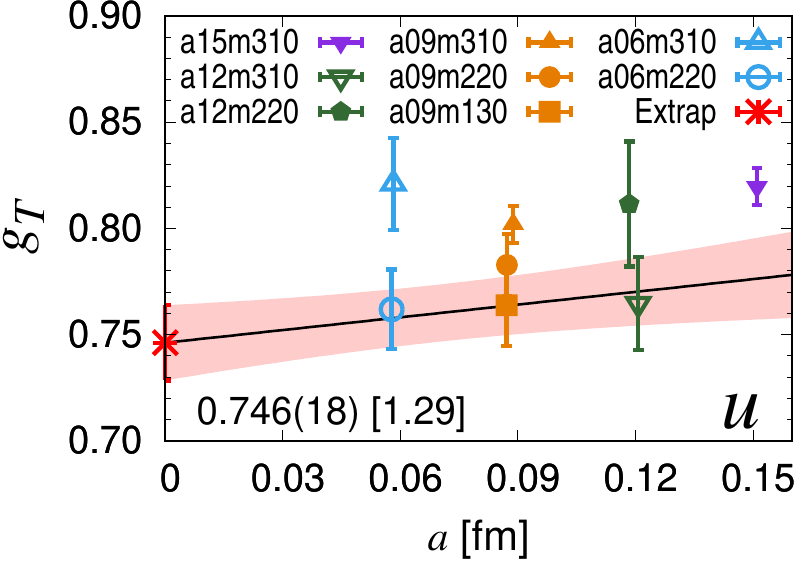}
      \includegraphics[width=0.235\linewidth]{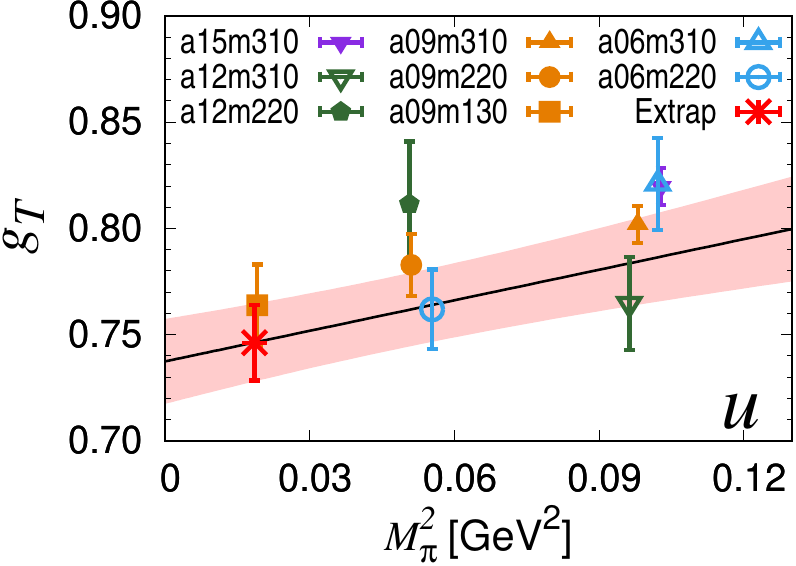}
  }\end{subfigure}
  \begin{subfigure}[$g_{T}^{d}$]{
      \includegraphics[width=0.235\linewidth]{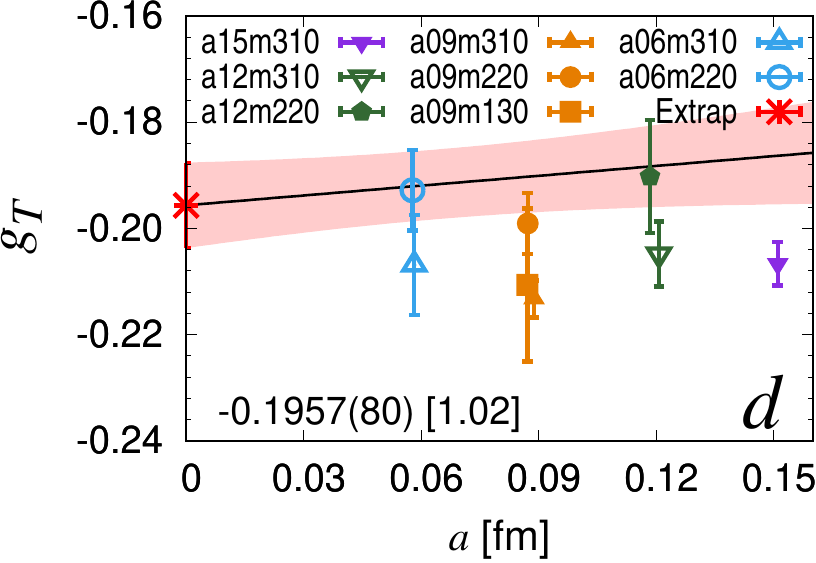}
      \includegraphics[width=0.235\linewidth]{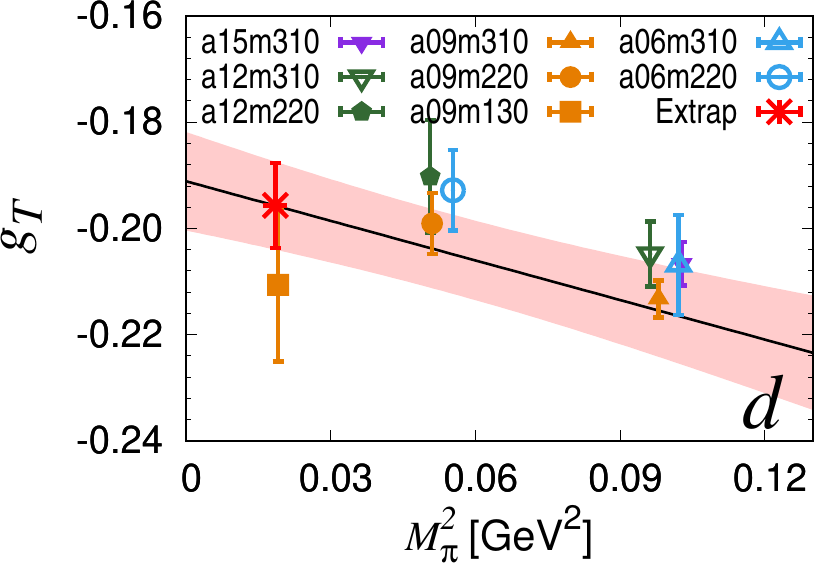}
  }\end{subfigure}
  \begin{subfigure}[$g_{T}^{u+d}$]{
      \includegraphics[width=0.235\linewidth]{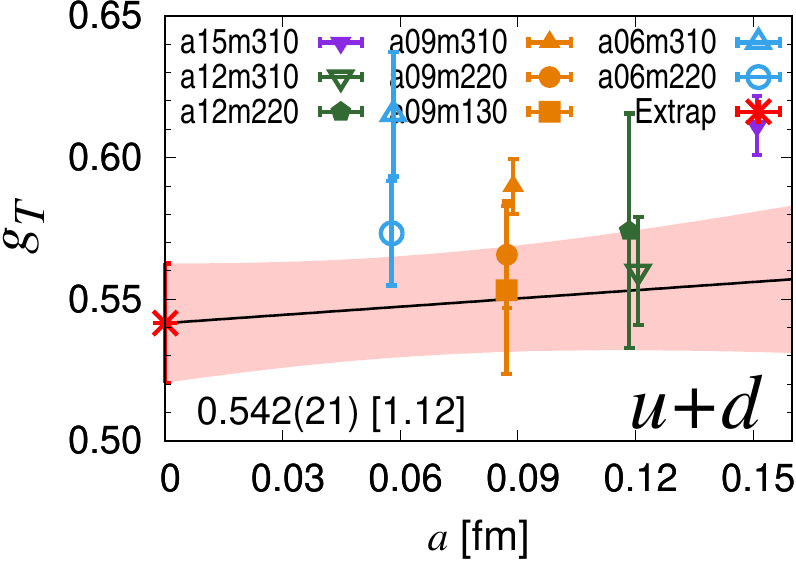}
      \includegraphics[width=0.235\linewidth]{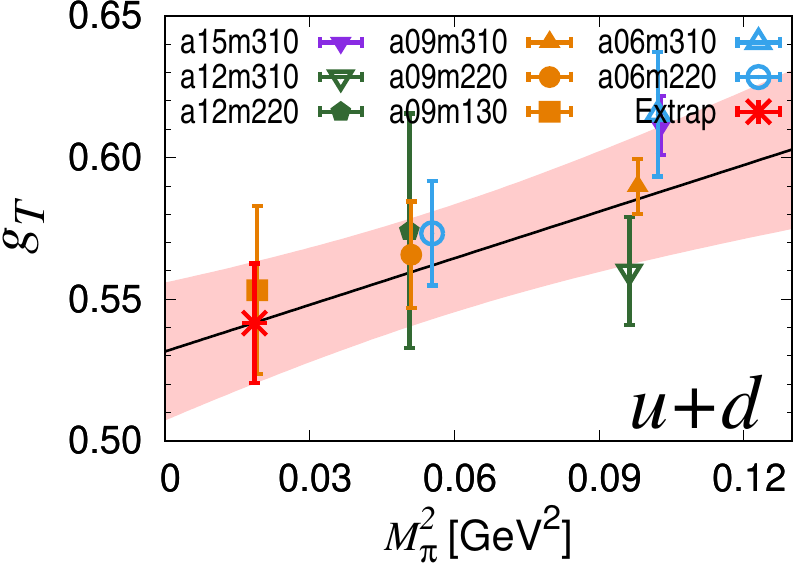}
  }\end{subfigure}
  \begin{subfigure}[$g_{T}^{s}$]{
      \includegraphics[width=0.235\linewidth]{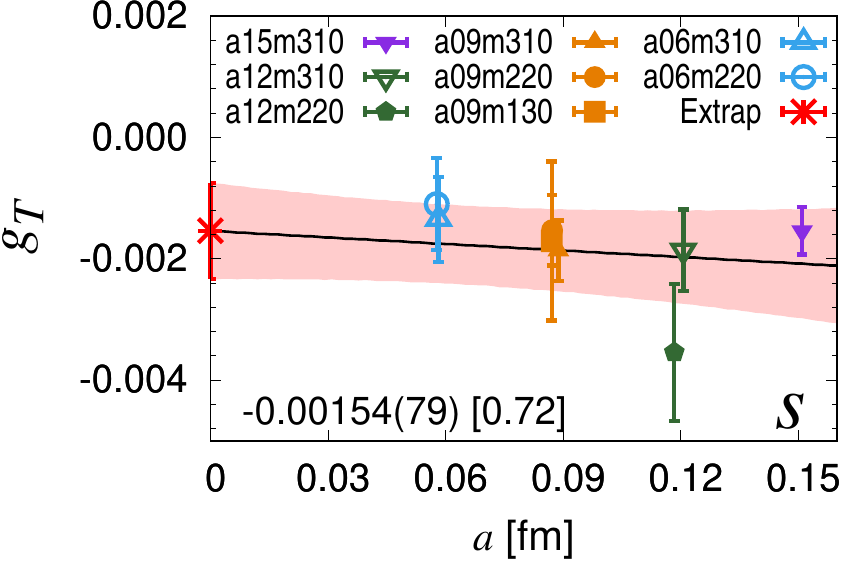}
      \includegraphics[width=0.235\linewidth]{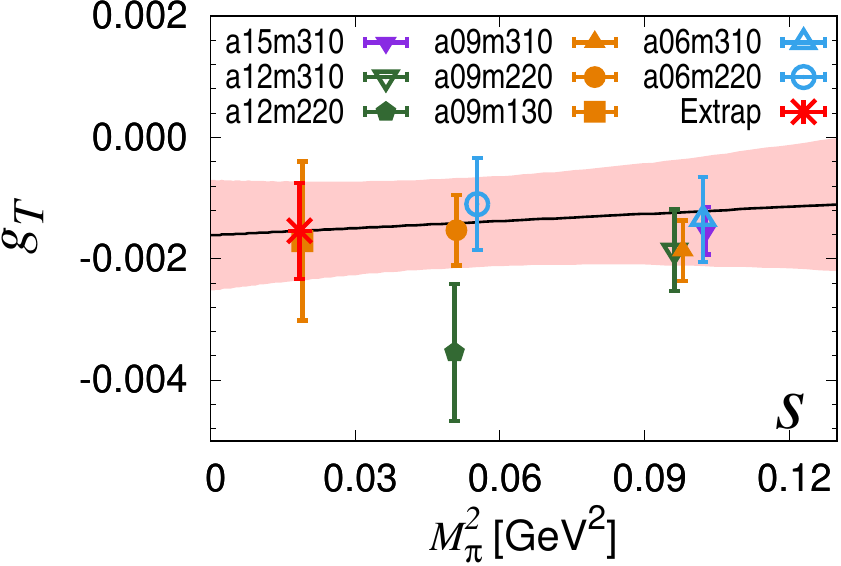}
  }\end{subfigure}
  \vspace{-0.1in}
  \caption{Chiral-continuum fits to $g_{T}$ using the ansatz $d_0+d_a a+ d_2 M_\pi^2$}
  \label{fig:CC_gT}
\end{figure}

\begin{figure}[p] 
  \centering

  \begin{subfigure}[$g_{S}^{u}$]{
      \includegraphics[width=0.235\linewidth]{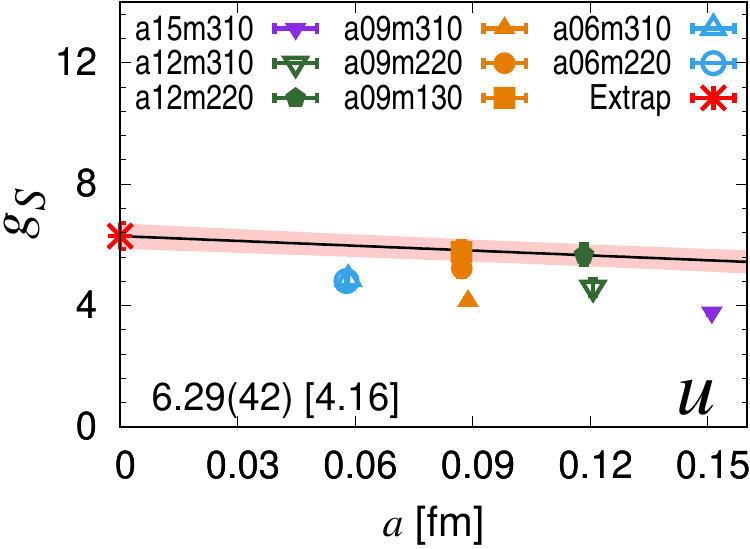}
      \includegraphics[width=0.235\linewidth]{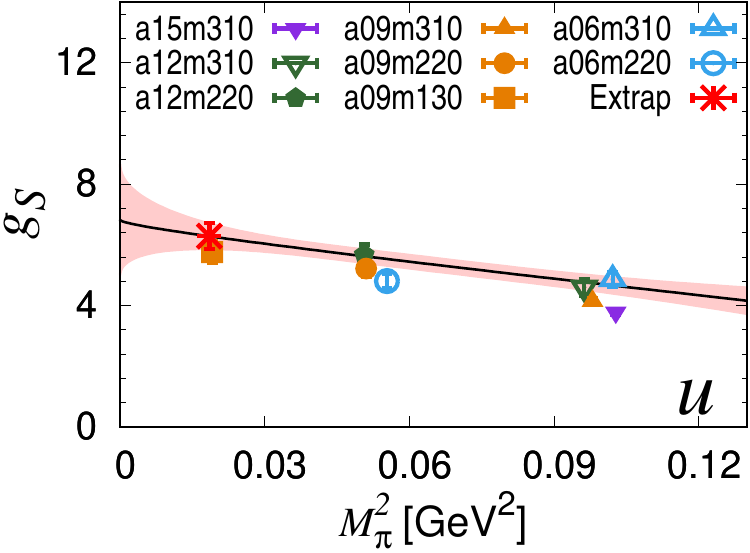}
  }\end{subfigure}
  \begin{subfigure}[$g_{S}^{d}$]{
      \includegraphics[width=0.235\linewidth]{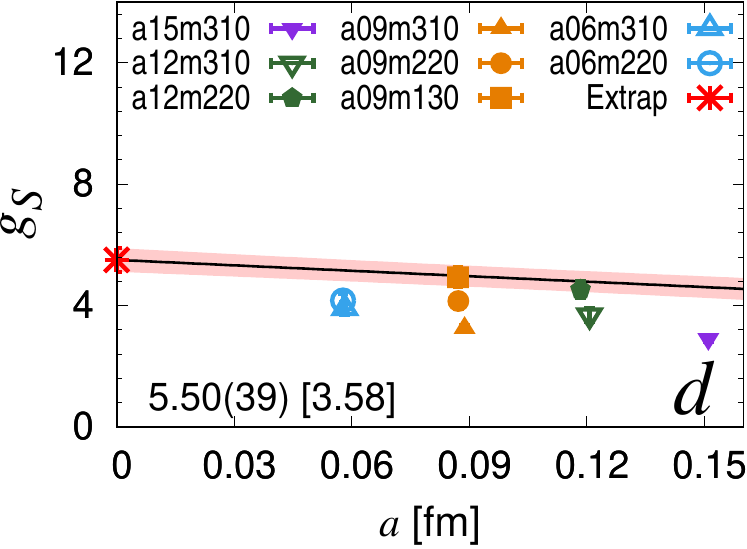}
      \includegraphics[width=0.235\linewidth]{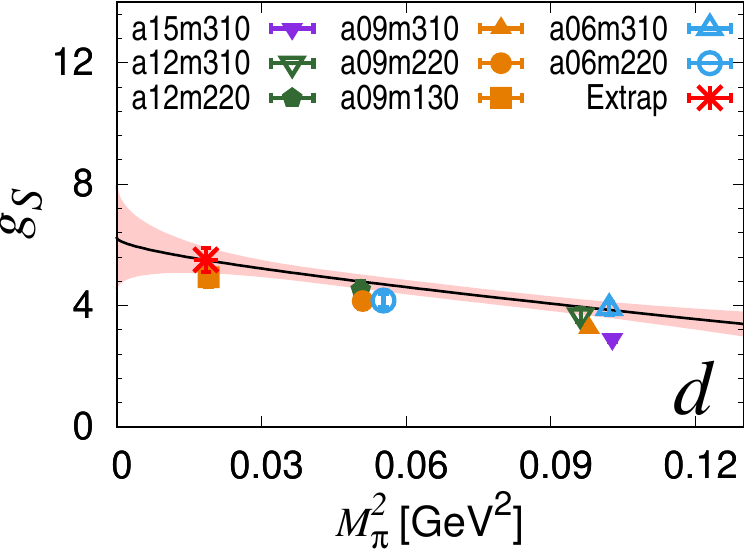}
  }\end{subfigure}
  \begin{subfigure}[$g_{S}^{u+d}$]{
      \includegraphics[width=0.235\linewidth]{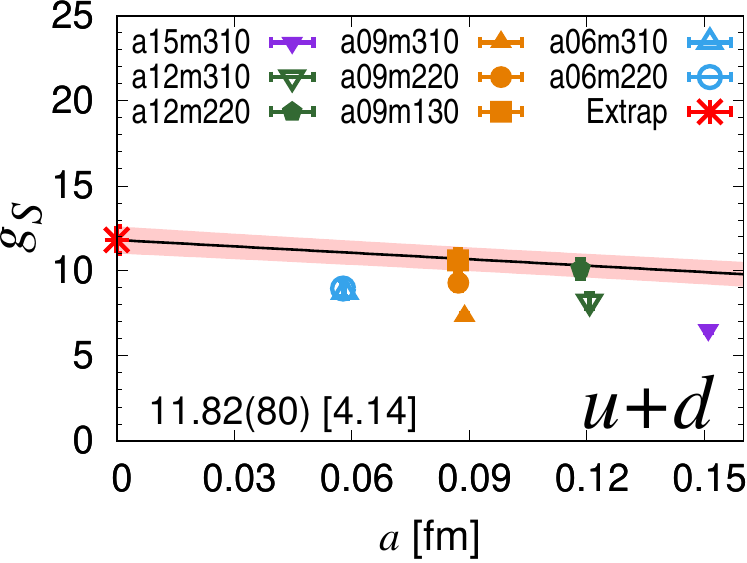}
      \includegraphics[width=0.235\linewidth]{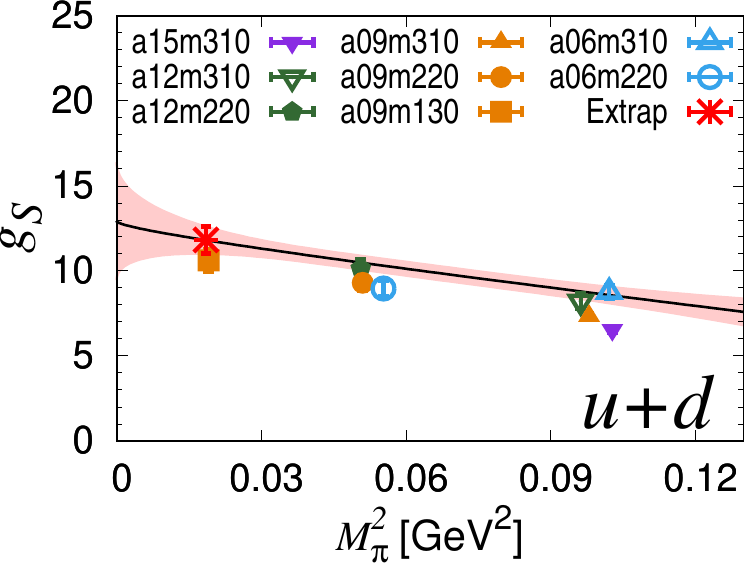}
  }\end{subfigure}
  \begin{subfigure}[$g_{S}^{s}$]{
      \includegraphics[width=0.235\linewidth]{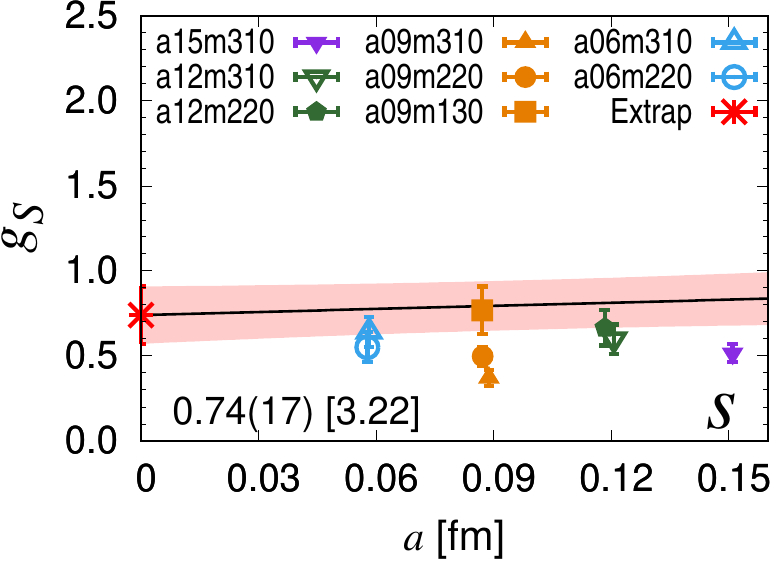}
      \includegraphics[width=0.235\linewidth]{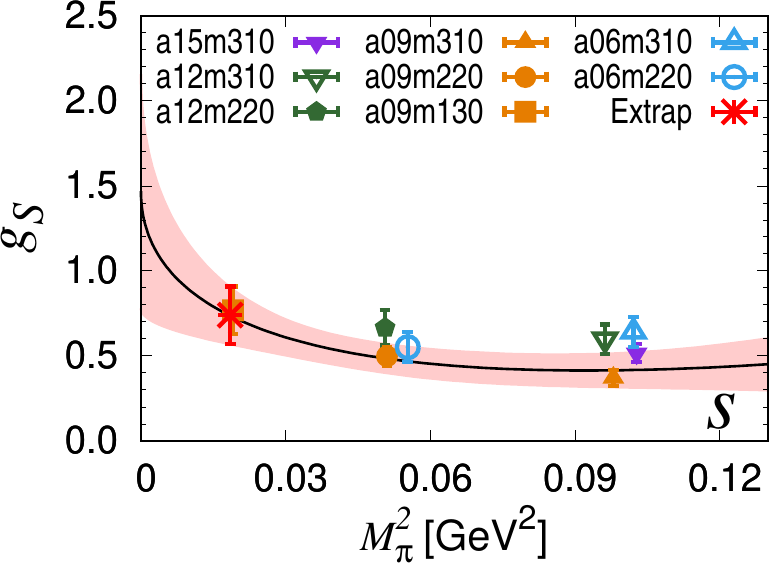}
  }\end{subfigure}
  \vspace{-0.1in}
  \caption{Chiral-continuum fits using the ansatz  $d_0+d_a a+ d_1 M_\pi+ d_2 M_\pi^2$ to $g_{S}$ from standard analysis.}
  \label{fig:CC_gS}
\end{figure}

\begin{figure}[p] 
  \centering
  
  \begin{subfigure}[$g_{S,N\pi}^{u}$]{
      \includegraphics[width=0.235\linewidth]{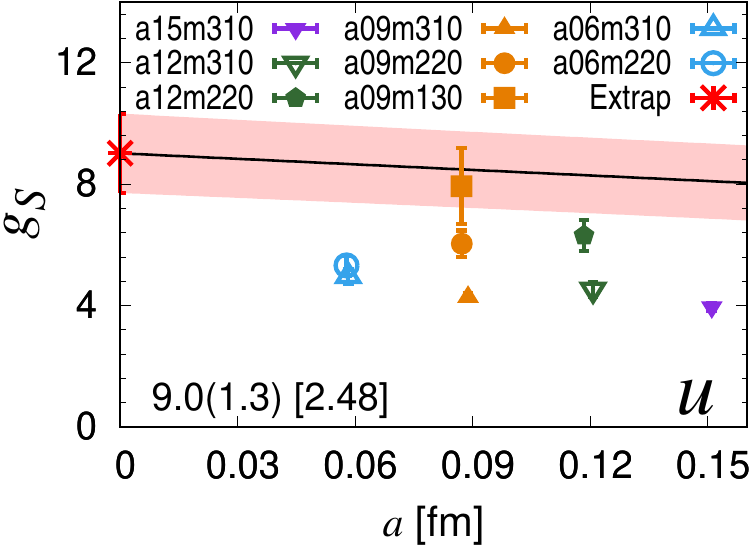}
      \includegraphics[width=0.235\linewidth]{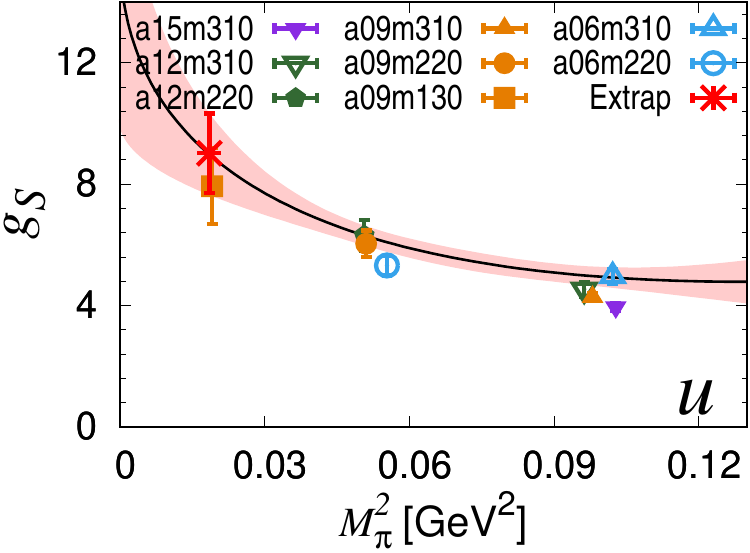}
  }\end{subfigure}
  \begin{subfigure}[$g_{S,N\pi}^{d}$]{
      \includegraphics[width=0.235\linewidth]{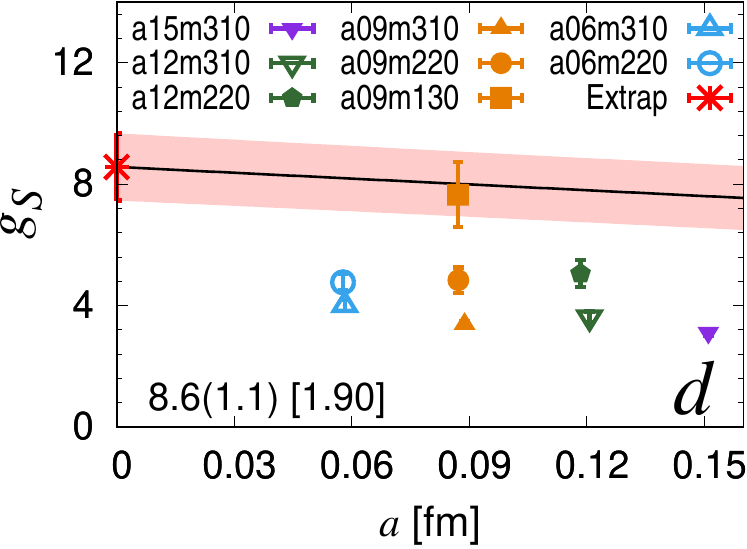}
      \includegraphics[width=0.235\linewidth]{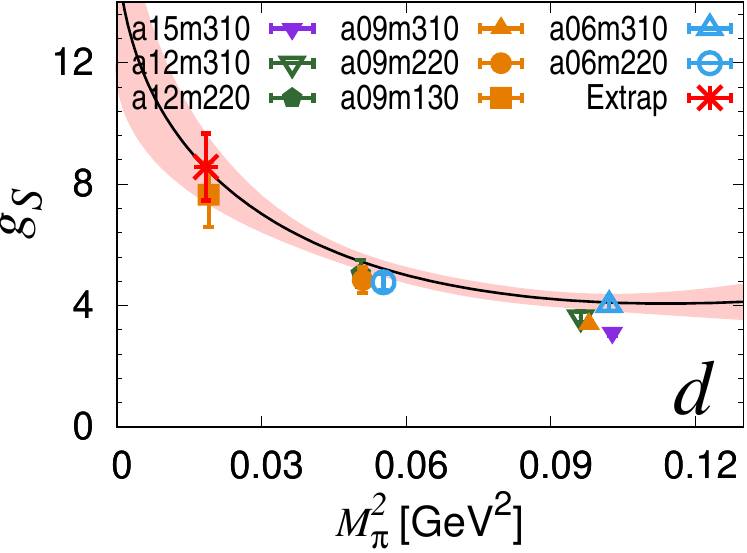}
  }\end{subfigure}
  \begin{subfigure}[$g_{S,N\pi}^{u+d}$]{
      \includegraphics[width=0.235\linewidth]{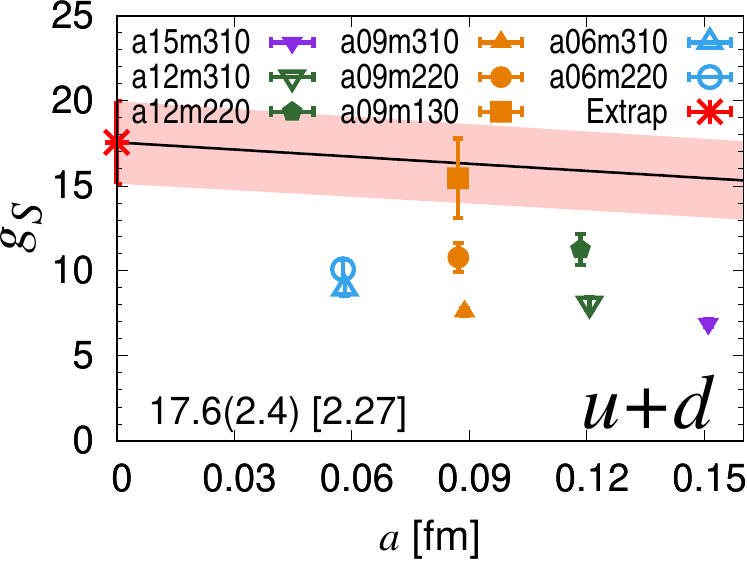}
      \includegraphics[width=0.235\linewidth]{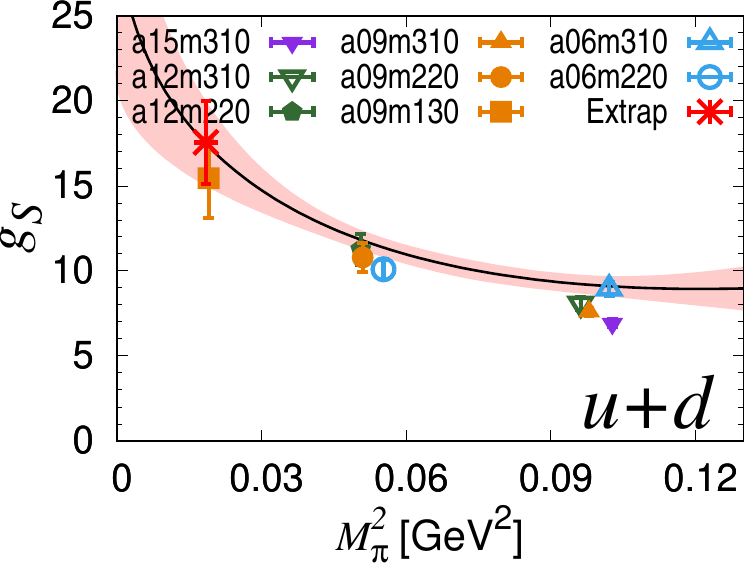}
  }\end{subfigure}
  \begin{subfigure}[$g_{S,N\pi}^{s}$]{
      \includegraphics[width=0.235\linewidth]{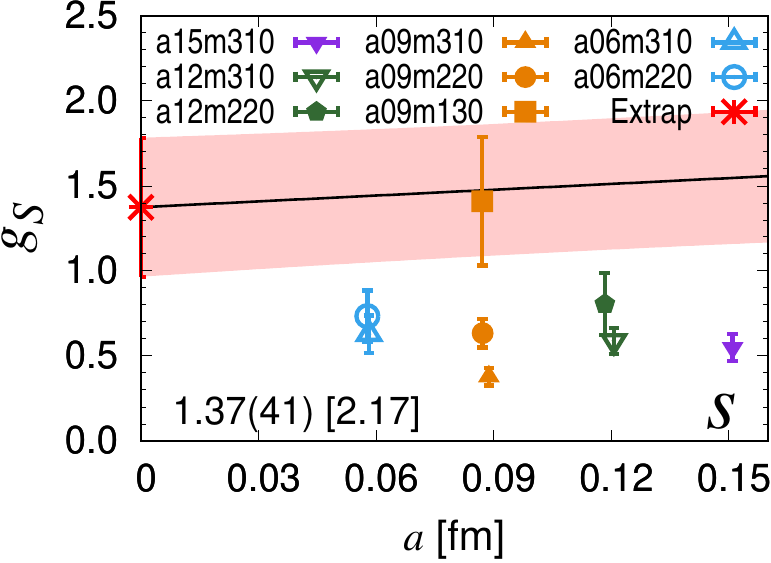}
      \includegraphics[width=0.235\linewidth]{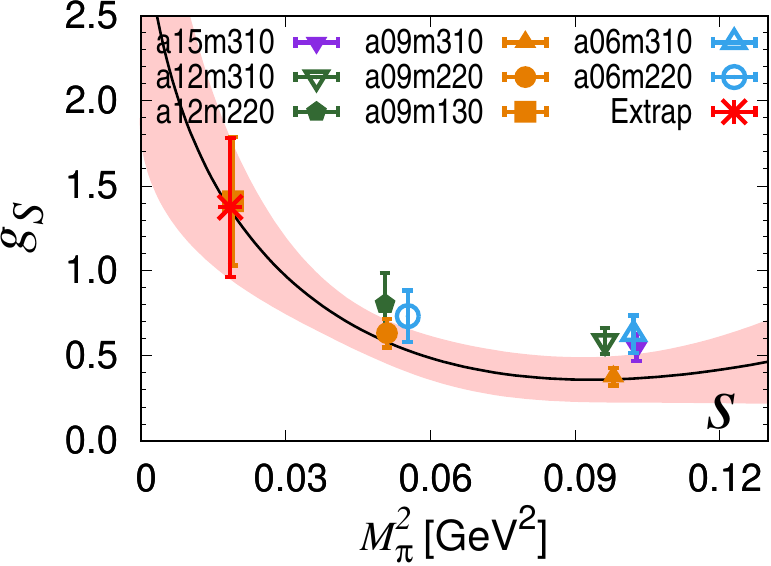}
  }\end{subfigure}
  \vspace{-0.1in}
  \caption{Chiral-continuum fits using the ansatz  $d_0+d_a a+ d_1 M_\pi+ d_2 M_\pi^2$ to $g_{S}$ from $N \pi$ analysis.}
  \label{fig:CC_gSNpi}
\end{figure}

The data for $C^{\text{conn}}_\Gamma(t;\tau) +
C^{\text{disc}}_\Gamma(t;\tau)$ are more noisy than
$C^{\text{conn}}_\Gamma(t;\tau)$, so we are only able to include
two states (ground plus one excited) when removing ESC using Eq.~\eqref{eq:3pt-sd}. In some
cases that show no obvious $\tau$ dependence, and the preliminary values
here are the unweighted average of central 5--6 points in the 3pt/2pt ratio
to get $g_T^s$.  Examples of ESC fits are shown
in Figs.~\ref{fig:gAgT} and~\ref{fig:gS}. The chiral-continuum (CC)
fits to the renormalized charges are shown in
Figs.~\ref{fig:CC_gA},~\ref{fig:CC_gT},~\ref{fig:CC_gS}
and~\ref{fig:CC_gSNpi}.  Possible finite-volume corrections are
ignored. The final results are summarized in
Tables~\ref{tab:gAgT},~\ref{tab:gS_u+d_CC} and~\ref{tab:gS}.  Some
details are as follows:

{\textbf{Axial charges, $g_A^{u,d,u+d,s}$}}:
We quote results of 2-state fit to the data from the ``standard'' analysis
of ESC.  The data are noisy and do not show clear $\tau$ dependence
for $q=u,s$ on the $a\approx 0.12$ and 0.15 fm ensembles.
The ESC in both $g_A^u$ and $g_A^d$ reduces $g_A^{u+d}$, with that in
$g_A^d$ being larger (see Figs.~\ref{fig:gAgT}).  Adding results of
separate fits to $g_A^u$ and $g_A^d$ gives values consistent with those from a single fit to
$g_A^{u+d}$.
The CC extrapolations are shown in
Fig.~\ref{fig:CC_gA}.  Both $g_A^u$ and $g_A^d$ show similar depencence on $a$ and
$\mpi$. There is a significant slope versus $M_\pi^2$.  The
$g_A^s$ data show a small dependence on $a$ and $\mpi$. The final
extrapolated $g_A^q$ are summarized in the Table
\ref{tab:gAgT}. Results for $g_A^{u,d}$ are consistent with those in
Ref.~\cite{Lin:2018obj}, while $g_A^s$ is $\approx 2\sigma$ smaller.

\begin{table}
  \centering
  \small
  \begin{tabular}{l | cccccc}
    & $g_A^{u}$ & $g_A^{d}$ & $g_A^{s}$ & $g_T^{u}$ & $g_T^{d}$ & $g_T^{s}$ \\
    \hline
    This work (preliminary) & 0.775(18) & -0.453(15) & -0.038(7) & 0.746(18) & -0.196(8) & -0.0015(8)\\
    PNDME 18~\protect\cite{Lin:2018obj,Gupta:2018lvp}  & 0.777(39) & -0.438(35) & -0.053(8) & 0.784(30) & -0.204(15) & -0.0032(7)\\
  \end{tabular}
\caption{Preliminary estimates of flavor diagonal axial and tensor charges compared 
  with our previous results, PNDME 18, from Refs.~\cite{Lin:2018obj,Gupta:2018lvp}. 
  The errors quoted under ``This work'' are only statistical.}
\label{tab:gAgT}
\end{table}

{\textbf{Tensor charges, $g_T^{u,d,u+d,s}$}}:
Quoted results are from the  ``standard'' analysis of ESC with 
2-state fit to $C^{\text{3pt},u}_T(t,\tau)$ and $C^{\text{3pt},d}_T(t,\tau)$ . 
The ESC in $g_T^u$ reduces while that in $g_T^d$ increases
$g_T^{u+d}$. The magnitude of ESC in $g_T^u$ is larger. Again,
combining results of separate fits to $g_T^u$ and $g_T^d$ gives values
consistent with results of fits to $g_T^{u+d}$.  As shown in
Fig.~\ref{fig:gT_s} and Ref.~\cite{Gupta:2018lvp}, there is no clear
ESC pattern in $C^{\text{3pt},s}_T(t;\tau)$, and the preliminary values
here are the unweighted average of central points in the 3pt/2pt ratio
to get $g_T^s$.
The CC fits are shown in Fig.~\ref{fig:CC_gT}, and the final extrapolated $g_T^q$ are
  summarized in the Table \ref{tab:gAgT}. Estimates of $g_T^{u,d}$ are consistent
  with Ref.~\cite{Gupta:2018lvp}, while again $g_T^s$ is smaller. \looseness-1



\begin{figure}[] 
  \includegraphics[width=0.24\linewidth]{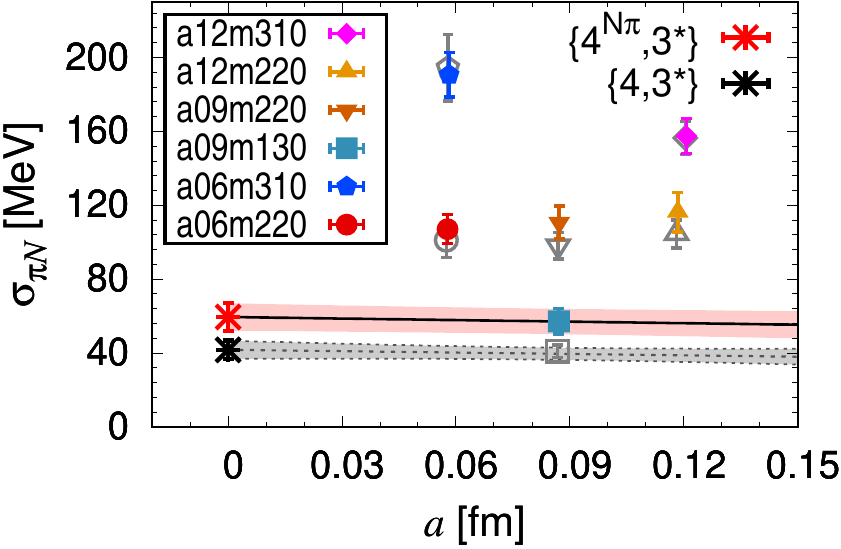}
  \includegraphics[width=0.24\linewidth]{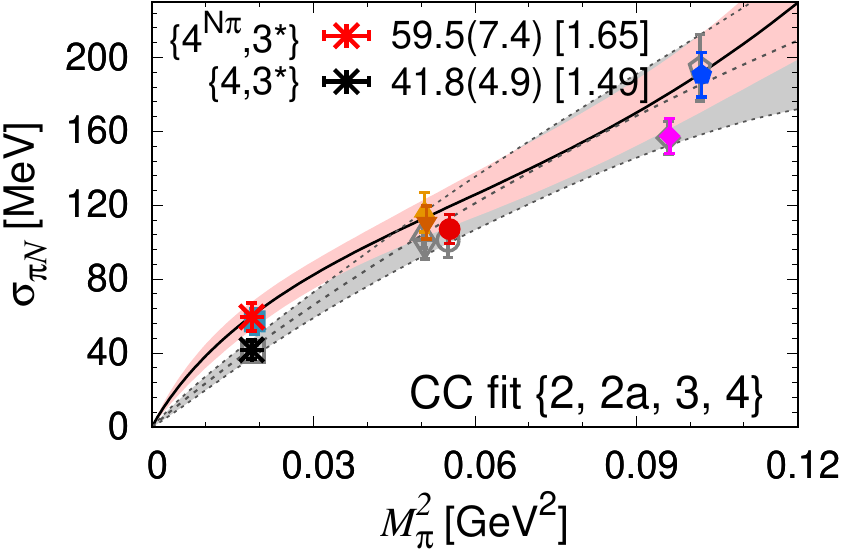}
  \includegraphics[width=0.24\linewidth]{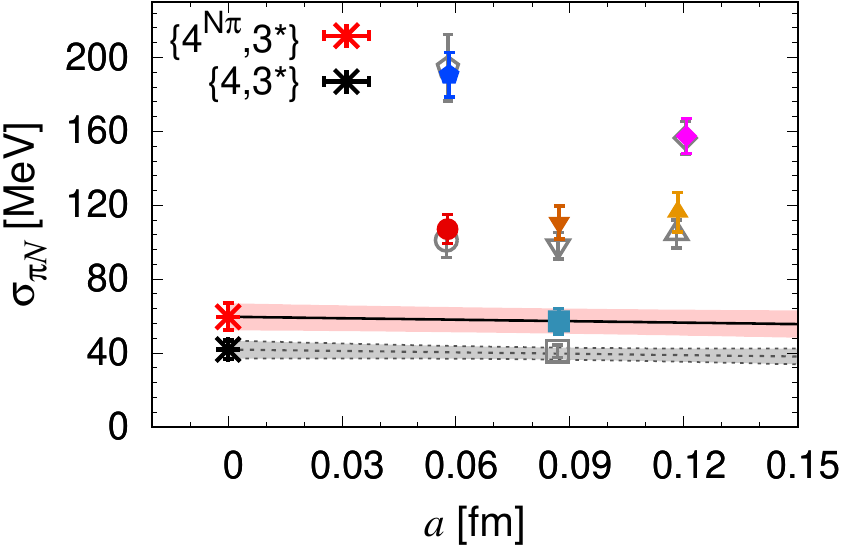}
  \includegraphics[width=0.24\linewidth]{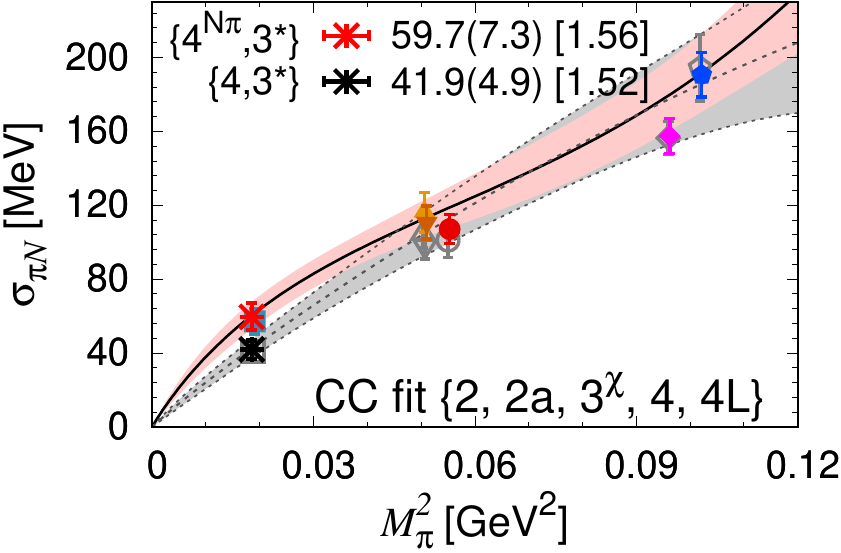}
  \caption{Data for the $\sigma$-term, $\sigma_{\pi N} = {m}_{ud}
    g_S^{u+d}$, from the two ESC strategies $\{4,3^\ast\}$ (gray) and
    $\{4^{N\pi},3^\ast\}$ (color) are shown as a function of $a$ and
    $\mpi^2$. These figures, reproduced from Ref.~\cite{Gupta:2021ahb}, 
    show CC fits $\{2,2a,3,4\}$ (left 2 panels) and $\{2,2a,3^\chi,4,4L\}$. 
    Final value of $\sigma_{\pi N}$ is given in the legends.}
    \label{fig:CCFV}
\end{figure}


{\textbf{Scalar charges $g_S^{u,d,u+d,s}$ and the pion-nucleon sigma term $\sigma_{\pi N}$}:} 
Chiral PT analyses provide two differences in the chiral behavior of flavor
diagonal scalar charges. First, the CC ansatz $d_0 + d_a a + d_1 \mpi
+ d_2 \mpi^2 + d_{2L} M_\pi^2 \log M_\pi^2$ has the chiral behavior
starting with a term proportional to
$\mpi$~\cite{Hoferichter:2015hva}. Second, the contribution of $N \pi$
and $N \pi \pi $ excited states is expected to be large in
$g_S^{u,d}$~\cite{Gupta:2021ahb}.  For $g_S^s$, the leading
multihadron excited state is expected to be $\Sigma K$, which has a
large mass gap, so we consider the ``standard'' analysis more
appropriate for it.
A comparison between using the ``standard'' and $N\pi$ strategies for
removing ESC in $g_S^{u,d}$ is shown in Fig.~\ref{fig:gS} for the
physical pion mass ensemble, $a09m130$. The ``$N\pi$'' analysis gives
a $40\sim50$\% larger value.  
The CC fits are shown in Figs.~\ref{fig:CC_gS} and \ref{fig:CC_gSNpi}.
To fit the expected chiral behavior in the ``$N\pi$'' analysis, with
at least 2 but likely more $M_\pi$-dependent terms contributing
significantly, requires data at 5--10 values of $M_\pi$. With data at
three values ($M_\pi \approx 135, 220, 310$~MeV), even 2 terms in the
fit ansatz is an overparametrization as is obvious from
Table~\ref{tab:gS_u+d_CC}. Our current estimates of $g_S^q$ are
summarized in Table~\ref{tab:gS}.

The analysis of $\sigma_{\pi N}$ has been presented recently in
Ref.~\cite{Gupta:2021ahb} and we reproduce the chiral fits from it in
Fig.~\ref{fig:CCFV} and the numbers in Table~\ref{tab:gS}.  The $N
\pi$ analysis gives $\sigma_{\pi N}|_{N\pi} \approx 60$~MeV,
consistent with phenomenology while the standard analysis gives a result 
consistent with previous lattice estimates, $\sigma_{\pi N}|_{\rm
  standard} \approx 40$~MeV~\cite{Gupta:2021ahb}. For $\sigma_s$, we recommend
using $g_S^s$ from ``without $N\pi$'' analyses. 

\begin{table} 
  \small
  \centering
  \begin{tabular}{l | cccccccc  }
    Fit ansatz & $d_0$ & $d_a$  & $d_1$ & $d_2$ & $d_{2L}$ & $\frac{\chi^2}{dof}$ & $g_S^{u+d}$
    \\ \hline
    $\{a,M_\pi,M_\pi^2\}$ &     $32(10)$ & $-14.0(4.2)$ &   $-131(77)$ &   $189(144)$ &           -- & $2.27$ & $17.6(2.4)$ \\
    $\{a,M_\pi,M_\pi^2,M_\pi^2\log M_\pi\}$ &  $0.7(48.9)$ & $-15.3(4.6)$ &   $447(891)$ &   $142(161)$ &   $638(979)$ & $2.89$ & $17.1(2.5)$ \\
    $\{a,M_\pi^2,M_\pi^2\log M_\pi\}$ &  $25.0(6.0)$ & $-14.3(4.2)$ &           -- &   $184(137)$ &    $149(84)$ & $2.23$ & $17.5(2.4)$ \\
  \end{tabular}
\caption{Chiral-Continuum fit coefficients for the renormalized
  $g_S^{u+d}$ using the ansatz $d_0 + d_a a + d_1 \mpi + d_2 \mpi^2 +
  d_{2L} M_\pi^2 \log M_\pi^2$. Data are from 
  excited states fits assuming the $N\pi$-state contributes.}
\label{tab:gS_u+d_CC}
\end{table}

\begin{table} 
  \centering
  \begin{tabular}{l | cccc}
    & $g_S^{u}$ & $g_S^{d}$ & $g_S^{s}$ & $\sigma_{\pi N}$ [MeV] \cite{Gupta:2021ahb}\\
    \hline
    without $N\pi$ & 6.29(42) & 5.50(39) & 0.74(17) & 41.9(4.9)\\
    with $N\pi$    & 9.0(1.3) & 8.6(1.1) & 1.37(41) & 59.6(7.4)\\
  \end{tabular}
  \caption{Flavor diagonal scalar charges. Results for the pion-nucleon sigma term $\sigma_{\pi N}$ are from Ref.~\cite{Gupta:2021ahb}. }
  \label{tab:gS}
\end{table}

{\textbf{Conclusion:}} Significant progress has been made in calculating $\chi$PT predictions
and using them in fits to remove ESC and do the chiral extrapolation.
Fits to present data do not, in most cases, 
distinguish between standard and $N\pi$ analyses. In the absence of
a new methodology, higher statistics data at many more values of $M_\pi$ and
on more physical pion mass ensembles are needed to get percent level
estimates for many quantities of phenomenological interest.



\section{Acknowledgements}
We thank the MILC collaboration for providing the 2+1+1-flavor HISQ
lattices. The calculations used the Chroma software
suite~\cite{Edwards:2004sx}.  This research used resources at (i) the
National Energy Research Scientific Computing Center, a DOE Office of
Science User Facility supported by the Office of Science of the
U.S.\ Department of Energy under Contract No.\ DE-AC02-05CH11231; (ii)
the Oak Ridge Leadership Computing Facility, which is a DOE Office of
Science User Facility supported under Contract DE-AC05-00OR22725, and
was awarded through the ALCC program project LGT107; (iii) the USQCD
collaboration, which is funded by the Office of Science of the
U.S.\ Department of Energy; and (iv) Institutional Computing at Los
Alamos National Laboratory. T.~Bhattacharya and R.~Gupta were partly
supported by the U.S.\ Department of Energy, Office of Science, Office
of High Energy Physics under Contract
No.\ DE-AC52-06NA25396. T.~Bhattacharya, R.~Gupta, E.~Mereghetti,
S.~Mondal, S.~Park, and B.~Yoon were partly supported by the LANL LDRD
program, and S.~Park by the Center for Nonlinear Studies.

\bibliographystyle{JHEP}
\bibliography{ref} 

\providecommand{\href}[2]{#2}\begingroup\raggedright\begin{thebibliography}{1}

\bibitem{Bazavov:2012xda}
{\bf MILC} Collaboration, A.~Bazavov {\em et~al.} {\em Phys. Rev.} {\bf D87}
  (2013), no.~5 054505, [\href{http://xxx.lanl.gov/abs/1212.4768}{{\tt
  1212.4768}}].

\bibitem{Gupta:2018qil}
R.~Gupta, Y.-C. Jang, B.~Yoon, H.-W. Lin, V.~Cirigliano, and T.~Bhattacharya
  {\em Phys. Rev.} {\bf D98} (2018) 034503,
  [\href{http://xxx.lanl.gov/abs/1806.09006}{{\tt 1806.09006}}].

\bibitem{Gupta:2018lvp}
R.~Gupta, B.~Yoon, T.~Bhattacharya, V.~Cirigliano, Y.-C. Jang, and H.-W. Lin
  {\em Phys. Rev.} {\bf D98} (2018), no.~9 091501,
  [\href{http://xxx.lanl.gov/abs/1808.07597}{{\tt 1808.07597}}].

\bibitem{Lin:2018obj}
H.-W. Lin, R.~Gupta, B.~Yoon, Y.-C. Jang, and T.~Bhattacharya {\em Phys. Rev.}
  {\bf D98} (2018), no.~9 094512,
  [\href{http://xxx.lanl.gov/abs/1806.10604}{{\tt 1806.10604}}].

\bibitem{Gupta:2021ahb}
R.~Gupta, S.~Park, M.~Hoferichter, E.~Mereghetti, B.~Yoon, and T.~Bhattacharya
  {\em Phys. Rev. Lett.} {\bf 127} (2021), no.~24 242002,
  [\href{http://xxx.lanl.gov/abs/2105.12095}{{\tt 2105.12095}}].

\bibitem{Bhattacharya:2015wna}
{\bf PNDME} Collaboration, T.~Bhattacharya, V.~Cirigliano, S.~Cohen, R.~Gupta,
  A.~Joseph, H.-W. Lin, and B.~Yoon {\em Phys. Rev.} {\bf D92} (2015), no.~9
  094511, [\href{http://xxx.lanl.gov/abs/1506.06411}{{\tt 1506.06411}}].

\bibitem{Park:2020axe}
S.~Park, T.~Bhattacharya, R.~Gupta, Y.-C. Jang, B.~Joo, H.-W. Lin, and B.~Yoon
  {\em PoS} {\bf LATTICE2019} (2020) 136,
  [\href{http://xxx.lanl.gov/abs/2002.02147}{{\tt 2002.02147}}].

\bibitem{Hoferichter:2015hva}
M.~Hoferichter, J.~Ruiz~de Elvira, B.~Kubis, and U.-G. Mei\ss{}ner {\em Phys.
  Rept.} {\bf 625} (2016) 1--88,
  [\href{http://xxx.lanl.gov/abs/1510.06039}{{\tt 1510.06039}}].

\bibitem{Edwards:2004sx}
{\bf SciDAC, LHPC, UKQCD} Collaboration, R.~G. Edwards and B.~Jo{\'o} {\em
  Nucl. Phys. Proc. Suppl.} {\bf 140} (2005) 832,
  [\href{http://xxx.lanl.gov/abs/hep-lat/0409003}{{\tt hep-lat/0409003}}].

\end{thebibliography}\endgroup

\end{document}